\documentclass[a4paper, 12pt]{article}
\usepackage{geometry}
\usepackage[skip=12pt, indent=18pt]{parskip}

\usepackage{amsthm,amsmath,latexsym,amssymb,amsfonts,amscd}
\usepackage{graphics,lscape,fancyhdr,array,stmaryrd,euscript,wrapfig}
\usepackage{empheq,wrapfig}
\usepackage{verbatim,slashed}
\numberwithin{equation}{section}
\usepackage{hyperref,setspace}
\usepackage{tikz-cd}
\usepackage{mathrsfs}
\usepackage[numbers,sort&compress]{natbib}
\usepackage[nottoc]{tocbibind}

\newcommand{\pl}{\partial}

\newcommand{\fud}[2]{{}^{#1}{}_{#2}\,}
\newcommand{\fdu}[2]{{}_{#1}{}^{#2}\,}

\newcommand{\besubeqs}{\begin{subequations}}
\newcommand{\esubeqs}{\end{subequations}}

\begin{document}
\pagenumbering{gobble}
\hfill
\vskip 0.01\textheight
\begin{center}
{\Large\bfseries 
Symmetric vs. chiral approaches \\ \vspace{5pt }to massive fields with spin }

\vspace{0.4cm}

\vskip 0.03\textheight
\renewcommand{\thefootnote}{\fnsymbol{footnote}}
William \textsc{Delplanque}, 
Evgeny \textsc{Skvortsov}\footnote{Research Associate of the Fund for Scientific Research -- FNRS, Belgium. e-mail: evgeny.skvortsov@umons.ac.be}\footnote{Also at Lebedev Institute of Physics.}
\renewcommand{\thefootnote}{\arabic{footnote}}
\setcounter{footnote}{0}
\vskip 0.03\textheight

{\em Service de Physique de l'Univers, Champs et Gravitation, \\ Universit\'e de Mons, 20 place du Parc, 7000 Mons, 
Belgium}

\end{center}

\vskip 0.02\textheight

\begin{abstract}
Massive higher spin fields are notoriously difficult to introduce interactions when they are described by symmetric (spin)-tensors. An alternative approach is to use chiral description that does not have unphysical longitudinal modes. For low spin fields we show that chiral and symmetric approaches can be related via a family of invertible change of variables (equivalent to parent actions), which should facilitate introduction of consistent interactions in the symmetric approach and help to control parity in the chiral one. We consider some examples of electromagnetic and gravitational interactions and their transmutations when going to the chiral formulation. An interesting feature of the relation is how second class constraints get eliminated while preserving Lorentz invariance.
\end{abstract}

\newpage
\tableofcontents
\newpage
\section{(Extended) Introduction}
\label{sec:}
\pagenumbering{arabic}
\setcounter{page}{1}
All particles, elementary or not, must fall into Wigner's classification \cite{Wigner:1939cj} and correspond to certain unitary irreducible representations of the spacetime symmetry group. Fixing the background spacetime be $4d$ Minkowski\footnote{See e.g. \cite{Bekaert:2006py,Basile:2016aen,Bekaert:2017khg} for reviews that also cover generalizations to higher dimensions, exotic cases like continuous spin and the cosmological constant. } and setting aside exotic cases we are left either with massive or massless particles with spin (helicity, for massless). The classification itself is only about free particles, while those to be observed have to interact. The list of options does not make any difference between low spin, $s\leq2$, and higher spin, $s>2$, particles, but the interactions do! Massless higher spin particles admit a very limited number of options for interactions, see e.g. \cite{Bekaert:2022poo} for an overview.\footnote{In $4d$ one can show that the smallest theory with massless higher spin fields is chiral higher spin gravity \cite{Metsaev:1991mt,Metsaev:1991nb,Ponomarev:2016lrm,Skvortsov:2018jea,Skvortsov:2020wtf} and its contractions \cite{Ponomarev:2017nrr,Krasnov:2021nsq}. } As different from the massless ones, massive higher spin fields are not subject to any strong no-go theorems, at least classically,\footnote{As is well-known, the presence of a just one massive higher spin particle leads to violation of tree level unitarity and one way to cure the problem is to have infinitely many fields of arbitrarily high spin, as in string theory. Also, there are some results along the line ``if one does something simple then it does not work'' for massive higher spins. } but consistent interactions are very difficult to construct.

The recent surge of interest in massive higher spin fields comes from two applications: (a) the discovery of gravitational waves encourages to develop efficient techniques to compute various observables for compact binaries, one approach being to model compact rotating objects as massive higher spin particles interacting with gravity, see e.g. \cite{Buonanno:2022pgc}; (b) search for theories of fundamental interactions seem to indicate that (massive or massless) higher spin states may be needed to shed more light on the quantum gravity problem. Little is known at present about theories with massive higher spin fields.

While the present paper deals with gravitational and electromagnetic/gauge non-Abelian interactions of massive fields with spins $s=\tfrac12,1,\tfrac32, 2$, most of the problems seem to be similar to the genuine higher spin case. Therefore, let us discuss the problem of higher spin interactions, where 'higher' starts from $s=\tfrac32$ for massive fields. 

In constructing consistent interactions of massive higher spin fields the main difficulty is to ensure that the unphysical components of the fields do not propagate. That the problem of massive higher spin fields is complicated can be observed already at the free level. Indeed, the most natural choice for a Lorentz covariant field to host massive spin-$s$ particle's degrees of freedom is to take a symmetric tensor field $\phi_{\mu_1\ldots \mu_s}$, hence, the \textit{symmetric} approach. The irreducibility constraints that need to be imposed on the field to put it on-shell and to project out unwanted (ghostly) components 
\begin{align}
    (\square -m^2)\phi_{\mu_1\ldots\mu_s}&=0\,, &
    \pl^\nu \phi_{\nu\mu_2\ldots\mu_s}&=0\,, &
    \phi\fud{\nu}{\nu\mu_3\ldots \mu_s}&=0\,,
\end{align}
are too many to admit a Lagrangian as they are. The last, the tracelessness, constraint can be imposed from the onset. Then, the first one admits a simple action. It is the second one, the transversality constraint that makes the system non-Lagrangian.\footnote{One can keep differential constraints on fields and gauge parameters at least for free fields, see e.g. \cite{Francia:2013sca,Abakumova:2020ajc}, but they tend to backfire when switching on interactions and quantizing the theory. }

One way out, proposed by Fierz and Pauli \cite{Fierz:1939ix}, which was elaborated on by Singh and Hagen \cite{Singh:1974qz,Singh:1974rc}, is to add a collection of auxiliary fields $\chi_k$ of ranks $k=s-2,s-3,\ldots 1,0$ and arrange the free action in such a way that the equations of motion imply various low-derivative consequences resulting eventually in the transversality constraint and ``suicide'' of 
auxiliary fields $\chi_k=0$. For the $s=2$ case one can combine $\phi_{\mu\nu}$ and $\chi_0$ into a traceful symmetric tensor. From the Dirac constraints analysis vantage point these actions lead to second class constraints. As different from the first class constraints that are usually manifested by gauge symmetries, the second class constraints are not immediately visible in the action. Therefore, the number of degrees of freedom is difficult to control. Most of the 'interaction terms' one can write down turn out to be inconsistent: some of the constraints get lost which increases the number of degrees of freedom, also known as Boulware-Deser ghost problem \cite{Boulware:1972yco}, whose particular solutions for massive spin-two fields are known as massive (bi)gravity \cite{Bergshoeff:2009hq,deRham:2010kj,Hassan:2011zd,deRham:2014zqa}.

A further development of the approach is to convert the second class constraints into the first class, thereby making it possible to control them via usual gauge symmetries of an action. This was achieved by Zinoviev for any (integer) spin $s$ in \cite{Zinoviev:2001dt}. The solution has a very suggestive form: an action for a massive spin-$s$ field is constructed as a sum of actions for massless fields with spins from $0$ to $s$ supplemented with one-derivative and no-derivative mixing terms. Indeed, the massive spin-$s$ particle's degrees of freedom are those of this sum of the massless ones. The low-derivative terms in the action make sure that we do not just have a direct sum of massless representations. A single massless spin-$k$ field requires traceless tensors of ranks $k$ and $k-2$ and has a gauge parameter that is a traceless rank-$(k-1)$ tensor. The two fields can be packaged into a double-traceless Fronsdal field \cite{Fronsdal:1978rb} $\Phi_k\equiv \Phi_{\mu_1 \ldots \mu_k}$, $\Phi\fud{\nu}{\nu \mu_3 \ldots \mu_k} \equiv0$. 
\begin{align}
\Phi_k&: &&\left\{\begin{aligned}
    &\boldsymbol{s} && s-1 && s-2 &&s-3 && \ldots && 2 && 1 &&0 \\
    &\boldsymbol{s-2} && \boldsymbol{s-3} && \boldsymbol{s-4} &&\boldsymbol{s-5} && \ldots && \boldsymbol{0} && \empty &&\empty 
\end{aligned}\right. \\
   \xi_k&: &&s-1 \quad s-2 \,\,\quad  s-3 \,\,\quad  s-4 \,\,\quad  \,\,\ldots \,\,\quad  1 \,\quad  0 &&\empty{} 
\end{align}
The gauge transformation has a very special structure: every parameter enters algebraically (shift symmetry) to the gauge transformation of the neighboring field
\begin{align}
    \delta \Phi_k &= \nabla \xi_{k} +\xi_{k+1} +g_{..} \xi_{k-1}\,,
\end{align}
where $g$ is the background metric ($(A)dS$ or flat). Therefore, one can simply eliminate as many fields as we have the gauge parameters. This leaves us with the boldface entries above, which is exactly the field content of the Singh-Hagen action. 

Now, interactions are easy to introduce, in principle: one should write the most general ansatz for interactions and gauge transformations and fix them by imposing gauge invariance, see e.g. \cite{Zinoviev:2006im,Zinoviev:2008ck,Zinoviev:2009hu,Zinoviev:2010cr,Buchbinder:2012iz}. The complications are of purely technical nature due to many tensor auxiliary fields being added to the theory. A generic feature of higher spin interactions is that the perturbative expansion does not stop at any finite order, except for low spin fields, and it is almost impossible to resum it without having a grasp on the underlying algebraic structures of the theory. Another aspect to be better understood are the constraints on the number of derivatives in vertices and gauge transformations.\footnote{Some restrictions on the number of derivatives are needed. Note that any expression in terms of $\Phi_s$ can be made gauge invariant, e.g. $A_\mu\rightarrow A_\mu-m^{-1} \pl_\mu \phi$ for spin-one. } Some other approaches to the problem of interactions of massive higher spin fields include covariant ideas \cite{Pashnev:1989gm,Buchbinder:2005ua,Bekaert:2003uc, Francia:2010qp,Buchbinder:2007ix,Buchbinder:2008ss,Kaparulin:2012px,Kazinski:2005eb,Metsaev:2012uy,Abakumova:2021evc,Abakumova:2023wve,Skvortsov:2023jbn} and the light-cone gauge \cite{Metsaev:2005ar,Metsaev:2007rn,Metsaev:2022yvb}.

An alternative approach to massive higher spins was proposed in \cite{Ochirov:2022nqz}, with low spin solutions along the same lines already available in the literature \cite{Chalmers:1997ui,Chalmers:2001cy} thanks to Chalmers and Siegel. The approach is confined to four spacetime dimensions. In $4d$ Lorentz (spin)-tensors are also representations of $sl(2,\mathbb{C})$. For example, a symmetric traceless tensor $\phi_{\mu_1 \ldots \mu_s}$ corresponds to $(s,s)$ of $sl(2,\mathbb{C})$, i.e. to $\phi_{A_1\ldots A_s, A'_1\ldots A_s'}$. The idea is that only $(2s,0)$ and $(0,2s)$ of $sl(2,\mathbb{C})$ contain the same number of (complex) components as the spin-$s$ representation of Wigner's little algebra $su(2)$, which is $2s+1$. Therefore, one can take $(2s,0)$, $\phi_{A_1\ldots A_{2s}}$, (or $(0,2s)$, but we stick to the first option) as the fundamental field variables that do not need any second class constraints to get down to the correct number of degrees of freedom. The field $\phi_{A_1\ldots A_{2s}}$ is both Lorentz-covariant and contains physical degrees of freedom only, but it is very chiral, hence, \textit{chiral} approach.

In the chiral approach there are no potentially dangerous unwanted components that have to be taken care of by transversality constraints. Interactions are very easy to introduce --- anything Lorentz invariant goes, but the price to pay is that parity/unitarity is not manifest. Typically, one needs certain non-minimal interactions to restore parity/unitarity. 

In order to control parity in the chiral approach and to build interactions in the symmetric approach in a more efficient way, it would be instructive to develop a map between the two approaches. It is not even clear a priori if such a map exists. The way to do it for $s=\tfrac12,1$ was already suggested by Chalmers and Siegel \cite{Chalmers:1997ui,Chalmers:2001cy}. We elaborate more on these simple cases as a warm-up and also because the gravitational interactions were not discussed in those papers. In the most general case one should be looking for a ``parent'' action that relates the Zinoviev/Singh-Hagen action to the chiral action via a series of steps of integrating in/out certain auxiliary fields. Given such a parent action relating any two theories, they are essentially equivalent, see e.g. \cite{Fradkin:1984ai}. 

We also consider the cases of $s=3/2,2$, which are new. The procedure of \textit{chiralization} takes one step for $s=3/2$ and two steps for $s=2$. The result gives a strong evidence that there exists parent action relating chiral and symmetric approaches for any spin. One genuine feature of the $s=2$ case is the step-by-step disappearance of the second class constraints in the process of chiralization without breaking the manifest Lorentz covariance. In a sense, passing to the chiral variables can be understood as a covariant way to parameterize the constraint surface. 

We also consider the transmutation of some interactions when going from the symmetric to the chiral approach. There are several types of electromagnetic/gauge non-Abelian and gravitational interactions one can try to introduce: a) all external fields are kept off-shell, which is the strongest possible option; b) the external fields can be put on-shell, i.e. satisfy Maxwell/Yang-Mills or Einstein equations; c) external fields can be kept small, i.e. expanded in powers of the field strength or Weyl tensor, or slow-varying, i.e. expanded in terms of the number of derivatives. We include interactions whenever possible for illustrative purposes and our ability to do that decays rapidly with spin, while the systematic study of various types of interactions is deferred to future papers. 

The outline of the rest of the paper is very simple: we study each of the four cases, $s=\tfrac12, 1, \tfrac32, 2$ one by one. 

\section{Spin-half}
\label{sec:half}
Let us begin with a toy-of-a-toy model illustrating the chiral formalism.\footnote{It is better to confess our notation early on. Indices $\mu,\nu,\ldots =0,\ldots,3$ are world indices on a (in general curved) spacetime manifold. We prefer the mostly plus signature. Indices $a,b,\ldots =0, \ldots, 3$ are those of the flat Minkowski or of the fiber space. Indices $A,B,\ldots = 1,2$ and $A',B',\ldots =1,2$ are indices of fundamental and anti-fundamental representations of the Lorentz algebra $sl(2,\mathbb{C})$. They are raised and lowered with the help of $\epsilon_{AB}$ and $\epsilon_{A'B'}$.  } The example is simple enough so that we can turn on gauge and/or gravitational interactions and the external fields can be kept off-shell. 

\paragraph{Chiralization of free fields.}
The Lagrangian density describing a free massive spin-half field $\psi_{A}$ in the ``symmetric approach''\footnote{It is a bit too early to call it symmetric since there is just one index.} is
\begin{align}
\mathcal{L} = \sqrt{2}\bar{\psi}^{A'}\partial_{AA'}\psi^{A} + \tfrac{1}{2}m\big(\psi^{A}\psi_{A} - \bar{\psi}^{A'}\bar{\psi}_{A'}\big) \, ,
\end{align}
where the coefficients are chosen in such a way that the Lagrangian density is Hermitian.\footnote{In the mostly plus signature $x^{AA'}$ is anti-Hermitian and so $\partial_{AA'}$ is. Classical fermionic fields are anti-commuting.} Here, $\psi_A$ and $\bar\psi_{A'}$ are in $(1,0)$ and $(0,1)$ representations of the Lorentz algebra $sl(2,\mathbb{C})$. It gives the following equations of motion
\begin{align}
\sqrt{2}\partial_{AA'}\bar{\psi}^{A'} + m\psi_{A} &= 0\,, \\
\sqrt{2}\partial_{AA'}\psi^{A} - m\bar{\psi}_{A'} &= 0 \, .
\end{align}
Interpreting the second equation as the definition of $\bar{\psi}_{A'}$ and plugging it into the first equation, we obtain 
\begin{align}
\big(\square-m^2\big)\psi_{A} = 0 \, ,
\end{align}
which is the Klein-Gordon equation describing the propagation of the two degrees of freedom of a massive spin-half field, characterized by the chiral field $\psi_{A}$. The Lagrangian density describing a free massive spin-half field in the chiral approach is therefore
\begin{align}
\mathcal{L} = \tfrac{1}{2}\psi^{A}\big(\square-m^2\big)\psi_{A} \, .
\end{align}

\paragraph{Chiralization of minimally interacting fields.} We will only consider the minimal Yang-Mills and gravitational interactions, i.e. the ones that are introduced via the covariant derivative 
\begin{align}
    \mathcal{D}&= d+ \tfrac 12 \omega^{a,b}\, T_{ab} +\mathcal{A}\,, && \mathcal{D}= dx^\mu\, e^{AA'}_\mu \mathcal{D}_{AA'}\,.
\end{align}
It contains both the minimal gravitational interaction via the spin-connection $\omega^{a,b}$ and the electromagnetic/Yang-Mills gauge field $\mathcal{A}\equiv \mathcal{A}_\mu dx^\mu$ (of some Lie algebra $\mathfrak{g}$, which we do not have to specify). We use $\nabla_{\! AA'}$ for the gravitational interaction and $D_{AA'}$ for the electromagnetic/Yang-Mills one. $e^{AA'}_\mu$ is a background vierbein. The spin-connection is solved for from the torsion constraint $\nabla e^{AA'}=0$. Generators $T_{ab}$ have been taught to act in the required representation of the fiber Lorentz algebra. The Yang-Mills algebra generators $T_I$ are always implicit, $\mathcal{A}=\mathcal{A}^{I} T_I$.  Given that we stick to the spinorial language it is more appropriate to write
\begin{align}
    \mathcal{D}&= d+ \tfrac 12 \omega^{AB}\, T_{AB}+ \tfrac 12 \omega^{A'B'}\, T_{A'B'} +\mathcal{A}\,, 
\end{align}
where $\omega^{AB}$, $\omega^{A'B'}$ are the (anti)-selfdual components of the spin-connection $\omega^{a,b}$ and $T_{AB}$, $T_{A'B'}$ are the Lorentz generators. The commutator of two covariant derivatives gives the curvature/field-strength
\begin{align}
[\mathcal{D}_{AA'},\mathcal{D}_{BB'}]\bullet := \mathcal{F}_{ABA'B'}\bullet \equiv \tfrac{1}{2}\epsilon_{A'B'}\mathcal{F}_{AB}\bullet + \tfrac{1}{2}\epsilon_{AB}\mathcal{F}_{A'B'}\bullet \, ,
\end{align}
where $\mathcal{F}_{AB} := \mathcal{F}_{ABC'}^{\phantom{ABC'}C'}$ and  $\mathcal{F}_{A'B'} := \mathcal{F}_{C\phantom{C}A'B'}^{\phantom{C}C}$. When two covariant derivatives are contracted, which often happens, we find
\besubeqs
\begin{align}
\mathcal{D}_{AA'}\mathcal{D}_{B}^{\phantom{B}A'}\bullet &\equiv \tfrac{1}{2}[\mathcal{D}_{AA'},\mathcal{D}_{B}^{\phantom{B}A'}]\bullet + \tfrac{1}{2}\{\mathcal{D}_{AA'},\mathcal{D}_{B}^{\phantom{B}A'}\}\bullet \equiv \tfrac{1}{2}\mathcal{F}_{AB}\bullet + \tfrac{1}{2}\epsilon_{AB}\square\bullet \,, \\
\mathcal{D}_{AA'}\mathcal{D}_{\phantom{A}B'}^{A}\bullet &\equiv \tfrac{1}{2}[\mathcal{D}_{AA'},\mathcal{D}_{\phantom{A}B'}^{A}]\bullet + \tfrac{1}{2}\{\mathcal{D}_{AA'},\mathcal{D}_{\phantom{A}B'}^{A}\}\bullet \equiv \tfrac{1}{2}\mathcal{F}_{A'B'}\bullet + \tfrac{1}{2}\epsilon_{A'B'}\square\bullet \, ,
\end{align}
\esubeqs
where $\square:=\mathcal{D}_{CC'}\mathcal{D}^{CC'}$. Finally, let us define our normalization for various components of the electromagnetic/Yang-Mills and gravitational field-strengths by acting with them on a spinor $\xi^A$ or $\xi^{A'}$ that can also be charged with respect to $\mathcal{A}$
\besubeqs
\begin{align}
\mathcal{F}_{AB}\xi^{C} &= F_{AB}\xi^{C} + R\epsilon_{A}^{\phantom{A}C}\xi_{B} + R\epsilon_{B}^{\phantom{B}C}\xi_{A} + C_{AB\phantom{C}D}^{\phantom{AB}C}\xi^{D} \, , \\
\mathcal{F}_{AB}\xi^{C'} &= F_{AB}\xi^{C'} + R_{AB\phantom{C'}D'}^{\phantom{AB}C'}\xi^{D'} \, , \\
\mathcal{F}_{A'B'}\xi^{C} &= F_{A'B'}\xi^{C} + R^{C}_{\phantom{C}DA'B'}\xi^{D} \, , \\
\mathcal{F}_{A'B'}\xi^{C'} &= F_{A'B'}\xi^{C'} + R\epsilon_{A'}^{\phantom{A'}C'}\xi_{B'} + R\epsilon_{B'}^{\phantom{B'}C'}\xi_{A'} + C_{A'B'\phantom{C'}D'}^{\phantom{A'B'}C'}\xi^{D'} \, ,
\end{align}
\esubeqs
where $F_{AB}$ is the field-strength of the electromagnetic/Yang-Mills interaction and $R$, $R_{ABA'B'}$ and $C_{ABCD}$, $C_{A'B'C'D'}$ are, respectively, the Ricci scalar, the traceless part of the Ricci tensor and the (anti)-selfdual components of the Weyl tensor.

The minimally coupled Dirac (Majorana) Lagrangian reads 
\begin{align}
\mathcal{L} = \sqrt{2}\bar{\psi}^{A'}\mathcal{D}_{AA'}\psi^{A} + \tfrac{1}{2}m\big(\psi^{A}\psi_{A} - \bar{\psi}^{A'}\bar{\psi}_{A'}\big) \, ,
\end{align}
and it gives the following equations of motion
\besubeqs
\begin{align}
\sqrt{2}\mathcal{D}_{AA'}\bar{\psi}^{A'} + m\psi_{A} &= 0\,, \\
\sqrt{2}\mathcal{D}_{AA'}\psi^{A} - m\bar{\psi}_{A'} &= 0 \, .
\end{align}
\esubeqs
Interpreting the second equation as the definition of $\bar{\psi}_{A'}$ and plugging it into the first one, we obtain an equation of motion for the chiral field
\begin{align}
\big(\square-m^2\big)\psi_{A} - \mathcal{F}_{AB}\psi^{B} = 0 \, .
\end{align}
Let us illustrate the key step of this simple calculation. After $\bar\psi_{A'}=\frac{\sqrt{2}}{m}\mathcal{D}_{AA'}\psi^A$ we find
\begin{align}
    \tfrac{2}{m} (\mathcal{D}_{AA'}\mathcal{D}\fdu{B}{A'})\psi^B+m\psi_A\equiv\tfrac{1}{m}\left(\mathcal{F}_{AB}\psi^B-\square \psi_A\right)+m\psi_A=0\,.
\end{align}
Let us note that the same manipulations can be performed at the level of the action (modulo total derivatives), but manipulations with the equations of motion are simpler due to them being linear in the fields. 
The Lagrangian density describing an interacting massive spin-half field in the chiral approach is therefore
\begin{align}
\mathcal{L} = \tfrac{1}{2}\psi^{A}\big(\square-m^2\big)\psi_{A} - \tfrac{1}{2}\psi^{A}\mathcal{F}_{AB}\psi^{B} \, ,
\end{align}
which differs by a non-minimal term from the naive chiral Lagrangian where only the minimal interaction, hidden in $\square$, is covariantized. In the case of electromagnetic/Yang-Mills interaction, we obtain
\begin{align}
\mathcal{L} = \tfrac{1}{2}\psi^{A}\big(\square-m^2\big)\psi_{A} - \tfrac{1}{2}\psi^{A}F_{AB}\psi^{B} \, .
\end{align}
Let us note that $\psi^A F_{AB} \psi^B$ implies the trace over the gauge group indices in the corresponding representation. For the gravitational interaction we find
\begin{align}
\mathcal{L} = \tfrac{1}{2}\psi^{A}\big(\square-m^2\big)\psi_{A} - \tfrac{3}{2}R\psi^{A}\psi_{A} \, .
\end{align}
The general lesson here is that the minimal interaction in the chiral approach needs to be supplemented by certain non-minimal terms for the theory to be parity-invariant. The non-minimal terms feature curvature couplings. 

\section{Spin-one}
\label{sec:}
For the Proca theory we begin with the case of free fields and then turn on gravitational and electromagnetic/Yang-Mills interactions. 

\subsection{Free fields: symmetric and chiral}
\label{sec:}
A free massive spin-one field can be described by the Proca action, whose Lagrangian reads
\begin{equation}
\mathcal{L} = -\tfrac{1}{4}F_{\mu\nu}F^{\mu\nu} - \tfrac{1}{2}m^2A_{\mu}A^{\mu} =-\tfrac{1}{2}\partial_{\mu}A_{\nu}\partial^{\mu}A^{\nu} + \tfrac{1}{2}\partial_{\mu}A_{\nu}\partial^{\nu}A^{\mu} - \tfrac{1}{2}m^2A_{\mu}A^{\mu}\,.
\end{equation}
The equation of motion 
\begin{align}
E_\mu&\equiv \square A_{\mu} - \partial_{\nu}\partial_{\mu}A^{\nu} - m^2 A_{\mu} = 0 
\end{align}
yield the transversality constraint upon taking the divergence
\begin{align}
\pl^\nu E_\nu= \partial_{\nu}A^{\nu} = 0  && \Longrightarrow && \big(\square - m^2\big)A_{\mu} = 0\,,
\end{align}
reducing thereby the number of degrees of freedom from four to three. As it was already discussed it is the transversality constraint that is difficult to maintain when interactions are turned on.  

In order to apply the procedure of chiralization, let us rewrite the Lagrangian density in the spinorial language as
\begin{align}
\mathcal{L} = -\tfrac{1}{2}\partial_{AA'}A_{BB'}\partial^{AA'}A^{BB'} + \tfrac{1}{2}\partial_{AA'}A_{BB'}\partial^{BB'}A^{AA'} - \tfrac{1}{2}m^2A_{AA'}A^{AA'} \, .
\end{align}
It gives the following equation of motion
\begin{align}
\square A_{AA'} - \partial_{BB'}\partial_{AA'}A^{BB'} - m^2 A_{AA'} = 0 \, .
\label{eq:spin1_free_EOM_(1,0)}
\end{align}
The first step of the procedure of chiralization is to define a new field, $\varphi_{AB}$, which is ``more chiral'' than $A_{BB'}$ as\footnote{The round brackets over the indices denote the symmetrization, which is defined to be a projector, i.e. one divides by the number of permutations. It is also convenient to denote all the indices to be symmetrized by the same letter, e.g. $V_A U_A\equiv \tfrac12 (V_{A_1} U_{A_2}+V_{A_2} U_{A_1})$. }
\begin{align}
\varphi_{AB} := m^{-1}\partial_{(A|A'|}^{\phantom{A}}A_{B)}^{\phantom{B)}A'} \, .
\label{eq:spin1_free_def_(2,0)}
\end{align}
By using this definition and the Fierz identity
\begin{align}
\Omega_{AB\phantom{B}\mathcal{X}}^{\phantom{AB}B} + \Omega_{B\phantom{B}A\mathcal{X}}^{\phantom{B}B} + \Omega^B_{\phantom{B}AB\mathcal{X}} \equiv 0 \,,
\label{Fierz}
\end{align}
where $\Omega$ is an arbitrary spin-tensor and $\mathcal{X}$ represents any other indices, we can show that the second-order equation of motion \eqref{eq:spin1_free_EOM_(1,0)} can be rewritten as the first-order one as
\begin{align}
A_{AA'} = 2m^{-1}\partial^{B}_{\phantom{B}A'}\varphi_{AB} \, .
\label{eq:spin1_free_(1,1)_fct_(2,0)}
\end{align}
Interpreting it as the definition of $A_{AA'}$ to replace it in the now-would-be equation for $\varphi_{AB}$ \eqref{eq:spin1_free_def_(2,0)}, we obtain 
\begin{align}
\big(\square-m^2\big)\varphi_{AB} = 0 \, ,
\label{eq:spin1_free_KG_(2,0)}
\end{align}
which is the desired Klein-Gordon equation, describing the free propagation of the three degrees of freedom of the massive spin-one field $\varphi_{AB}$. Therefore, the Lagrangian density in the chiral approach is simply
\begin{align}\label{chiralone}
\mathcal{L} = \tfrac{1}{2}\varphi^{AB}\big(\square-m^2\big)\varphi_{AB} \, .
\end{align}

\paragraph{What happened to the transversality constraint?} It is an interesting question! Let us take the expression \eqref{eq:spin1_free_(1,1)_fct_(2,0)} for $A_{BB'}$ in terms of $\varphi_{AB}$ and plug it into the transversality constraint $\partial_{AA'}A^{AA'} = 0 $, we find
\begin{align}
\partial_{AA'}A^{AA'} = \epsilon_{AB}\square\varphi^{AB} \equiv 0 \, ,
\end{align}
i.e. it is trivially satisfied. It means that the new set of ``coordinates'' $\varphi_{AB}$ satisfying just the Klein-Gordon equation correctly parameterizes the solution space of the Proca theory in terms of the old coordinates $A_{BB'}$.

\paragraph{Chiralization at the action level.} We performed the chiralization at the level of equations motion. It might be instructive to see how it works at the action level \cite{Chalmers:1997ui,Chalmers:2001cy}. The field strength decomposes as
\begin{equation}
F_{\mu\nu} \equiv F_{AA'BB'} = \tfrac{1}{2}F_{AB}\epsilon_{A'B'} + \tfrac{1}{2}F_{A'B'}\epsilon_{AB} \,,
\end{equation}
where 
\begin{equation}
F_{AB} := 2\partial_{(AC'}^{\phantom{A}}A_{B)}^{\phantom{B)}C'} \;\;\; \text{ and } \;\;\; F_{A'B'} := 2\partial_{(A'C}^{\phantom{A}}A_{B')}^{\phantom{B')}C} \,,
\end{equation}
Taking into account that $F_{\mu\nu}F^{\mu\nu}= \tfrac{1}{2}\big(F_{AB}F^{AB}+ F_{A'B'}F^{A'B'}\big)$ and
\begin{align}
    \int F\wedge F=\int d^4x\, \Big(F_{AB}F^{AB}- F_{A'B'}F^{A'B'}  \Big)  
\end{align}
is a topological invariant, we can eliminate $F_{A'B'}F^{A'B'}$ from the action to get 
\begin{align}
S[A] &= \int  \mathrm{d}^4 x\, \Big(-\tfrac{1}{4}F_{AB}F^{AB} - \tfrac{1}{2}m^2 A_{AA'}A^{AA'} \Big) \,.
\label{eq:Proca_action_spinor}
\end{align}
As different from the spin-half example, we do not have any chiral field in the initial action, just $A_{BB'}$. Chiral field $\varphi_{AB}$ can be integrated in since the first term of the action is a perfect square (we do not rescale $\varphi_{AB}$ by $m$, so it has a different dimension here):
\begin{equation}\label{Aphiaction}
S[A,\phi] = \int  \mathrm{d}^4 x\, \Big(-\tfrac{1}{2}\varphi_{AB}F^{AB} + \tfrac{1}{4}\varphi_{AB}\varphi^{AB} - \tfrac{1}{2}m^2 A_{AA'}A^{AA'} \Big) \text{ .}
\end{equation}
We have now two fields: $\varphi_{AB}$ and $A_{AA'}$. The equation of motion $\varphi_{AB} = F_{AB}$ for $\varphi_{AB}$ just defines it to be an avatar for $F_{AB}$. The equation for $A_{AA'}$ can be used now to solve for it
\begin{equation}
A_{AA'} = \frac{1}{m^2}\partial^{C}_{\phantom{C}A'}\varphi_{CA} \,.
\end{equation}
Using this last relation in the action \eqref{Aphiaction}, we obtain the Klein-Gordon action, \eqref{chiralone}, up to a rescaling of the field.

\subsubsection{Hamiltonian analysis}
\label{sec:}
Let us recall that from the Hamiltonian point of view Proca theory features second class constraints:
\begin{align}
    \chi_1&= p^0 &&\chi_2= \pl_i p^i -m^2 A^0 \, .
\end{align}
The constraints are not easily visible at the action level. Besides that, we have to break Lorentz symmetry to see them. Another idea is to manifest them via Stuckelberg gauge symmetries 
\begin{align}
    -\tfrac{1}{4}F_{\mu\nu}F^{\mu\nu} - \tfrac{1}{2}m^2(A_{\mu}-\pl_\mu \phi)(A^{\mu}-\pl^\mu \phi)\,,
\end{align}
where the action is invariant under $\delta A_\mu=\pl_\mu \xi$ and $\delta \phi=-\xi$.\footnote{This example is somewhat trivial since the Stueckelberg field $\phi$ can be introduced via a shift, which can always be done. For higher spin fields such shifts usually lead to higher derivative interactions for Stueckelberg fields, which, in general, is the sign of new unphysical degrees of freedom having been activated. } Via the Stueckelberg trick the second class constraints get converted into the first class ones. The latter are explicitly visible via the gauge parameters. Now the counting degrees of freedom is very simple: $\xi$ manifests a (primary) first class constraint and there is a secondary one. Usually, the number of first class constraints is just twice the number of gauge parameters. Therefore, $\#$fields $-$ $\#$ first class constraints, which gives $4(A)+1(\phi)-2\times 1(\xi)=3$. The latter does not require the $3+1$ split and Hamiltonian analysis. 

What happens when interactions are not introduced in a consistent way is that new (ghostly) degrees of freedom get activated. Within the Hamiltonian formalism this can be detected as a loss of some of the second class constraints, which is a tedious analysis for a nonlinear theory. Within the Stueckelberg formulation one just needs to maintain gauge invariance (provided certain constraints on the number of derivatives are imposed). 

An interesting feature of the chiralization is that the second class constraints are gone in the $\varphi_{AB}$ theory. In some sense, via integrating in and out some fields there is a way to ``solve'' the constraints as to reduce everything to the physical variables. The procedure does not break Lorentz invariance. It is also interesting to compare the parent action \eqref{Aphiaction} with a closely related first-order formulation of Proca theory, see Appendix \eqref{app:Proca}, the conclusion being is that the usual first-order formulation does not help to get rid off the second-class constraints in a covariant way.

\subsection{Gravitational interactions}
\label{sec:}
Gravitational interactions are somewhat easier to introduce. To study interactions we begin with the minimally coupled Proca Lagrangian
\begin{align}
\mathcal{L} = -\tfrac{1}{2}\mathcal{D}_{AA'}A_{BB'}\mathcal{D}^{AA'}A^{BB'} + \tfrac{1}{2}\mathcal{D}_{AA'}A_{BB'}\mathcal{D}^{BB'}A^{AA'} - \tfrac{1}{2}m^2A_{AA'}A^{AA'} \, ,
\end{align}
which gives the following equation of motion
\begin{align}
\square A_{AA'} - \mathcal{D}_{BB'}\mathcal{D}_{AA'}A^{BB'} - m^2 A_{AA'} = 0 \, .
\label{eq:spin1_gravit_EOM_(1,0)}
\end{align}
By taking the divergence of equation of motion, we find
\begin{align}
\mathcal{D}^{AA'}\mathcal{D}_{BB'}\mathcal{D}^{BB'}A_{AA'} - \mathcal{D}^{AA'}\mathcal{D}_{BB'}\mathcal{D}_{AA'}A^{BB'} - m^2 \mathcal{D}^{AA'}A_{AA'} &= 0 \\
\Leftrightarrow \quad - \tfrac{1}{2}\mathcal{F}_{AB}\mathcal{D}^{A}_{\phantom{A}B'}A^{BB'} - \tfrac{1}{2}\mathcal{F}_{A'B'}\mathcal{D}_{B}^{\phantom{B}A'}A^{BB'} - m^2 \mathcal{D}^{AA'}A_{AA'} &= 0 \, .
\label{eq:spin1_constraint_general_interaction}
\end{align}
The terms with the curvature in the last line simply vanish for the gravitational interaction. Indeed, in the vector language we have 
\begin{align}
    \mathcal{F}_{mn} (\mathcal{D}^n A^m)&= R_{mn} (\nabla^n A^m-\nabla^m A^n)\equiv0\,.
\end{align}
The difference between electromagnetic/Yang-Mills and gravitational interactions is that $\mathcal{F}_{AB}(G^{AB})$ is equal to $\rho(F_{AB})G^{AB}\neq0$ for the former and $0$ for the latter (here, $\rho$ is a representation of the gauge algebra where $G^{AB}=\mathcal{D}^{(A}_{\phantom{A}B'}A^{B)B'}_{\phantom{A}}$ takes values).

Therefore, in the case of gravitational interaction we preserve the transversality constraint without any further ado
\begin{align}
\nabla^{AA'}A_{AA'} = 0 \, .
\label{eq:spin1_grav_transverse_constraint}
\end{align}

\subsubsection{Chiralization}
For the procedure of chiralization we define the chiral field
\begin{align}
\varphi_{AB} := m^{-1}\nabla_{\!(A|A'|}^{\phantom{A}
}A_{B)}^{\phantom{B)}A'} \, .
\label{eq:spin1_gravit_def_(2,0)}
\end{align}
By using this definition, the Fierz identity and the commutators of covariant derivatives, we can rewrite the second-order equation of motion \eqref{eq:spin1_gravit_EOM_(1,0)} as the first-order one for $\varphi$\footnote{We keep the notation $\mathcal{F}$ for the commutator of covariant derivatives to have more compact expressions, but keep in mind that we treat the gravitational interaction here.}
\begin{align}
2m\nabla^{B}_{\phantom{B}A'}\varphi_{AB} - \tfrac{1}{2}\Big(\mathcal{F}_{AB}A^{B}_{\phantom{B}A'} - \mathcal{F}_{A'B'}A_{A}^{\phantom{A}B'}\Big) - m^2A_{AA'} = 0 \, .
\label{eq:spin1_interaction_EOM_(2,0)_(1,1)}
\end{align}
It is easy to show that
\begin{align}
\mathcal{F}_{AB}A^{B}_{\phantom{B}A'} - \mathcal{F}_{A'B'}A_{A}^{\phantom{A}B'} = 0
\end{align}
in the case of the gravitational interaction. Therefore, we can use the previous relation as the definition of the field $A_{AA'}$
\begin{align}
A_{AA'} = 2m^{-1}\nabla^{B}_{\phantom{B}A'}\varphi_{AB} \, .
\label{eq:spin1_grav_(1,1)_fct_(2,0)}
\end{align}
Plugging this definition into the now-would-be-equation for $\varphi_{AB}$ \eqref{eq:spin1_gravit_def_(2,0)}, we obtain the following second-order equation of motion
\begin{align}
\big(\square-m^2\big)\varphi_{AB} - 4R\varphi_{AB} - C_{ABCD}\varphi^{CD} = 0 \, .
\end{align}
Therefore, the Lagrangian density describing a massive spin-one field interacting with gravity in the chiral approach is 
\begin{align}
\mathcal{L} = \tfrac{1}{2}\varphi^{AB}\big(\square-m^2\big)\varphi_{AB} - 2R\varphi^{AB}\varphi_{AB} - \tfrac{1}{2}C_{ABCD}\varphi^{AB}\varphi^{CD} \, .
\end{align}
Again, we observe that in the chiral approach we need to add nonminimal interactions to the Lagrangian in order to get a parity invariant theory.

\paragraph{What happened to the transversality constraint?} It is again interesting to see whether in the chiral formulation we are consistent with the transversality constraint. We take $
A_{AA'} = 2m^{-1}\nabla^{B}_{\phantom{B}A'}\varphi_{AB} $ and plug it into \eqref{eq:spin1_grav_transverse_constraint} to get
\begin{align}
\nabla^{AA'}A_{AA'}=\epsilon_{AB}\square\varphi^{AB} + 2R\epsilon_{(A}^{\phantom{(A}A}\varphi_{B)}^{\phantom{B)}B} + 2R\epsilon_{(A}^{\phantom{(A}B}\varphi^{A}_{\phantom{A}B)} + C_{AB\phantom{A}C}^{\phantom{AB}A}\varphi^{CB} + C_{AB\phantom{A}C}^{\phantom{AB}B}\varphi^{AC} \equiv 0 \, ,
\end{align}
i.e. it is trivially satisfied. Therefore, $\varphi_{AB}$ are new but equivalent coordinates to parameterize the solution space.

\subsection{Electromagnetic/Yang-Mills interactions}
\label{sec:}
Let us also consider a model where a massive spin-one field is coupled to an electromagnetic field, e.g. $W$-bosons. A simplified version is without the $Z$-boson where the initial gauge field takes values in $so(3)$ that gets broken down to $u(1)$, see e.g. \cite{Cangemi:2023ysz} (or any QFT textbook). After the Higgs mechanism is realized we can thank the Higgs field for its service and decouple it while also imposing the unitary gauge to find
\begin{align}\label{completeh}
{\cal L} = -\tfrac{1}{4} (F_{\mu\nu})^2
- \tfrac{1}{2} |W_{\mu\nu}|^2+ m| W_\mu|^2-i q Q \overline W^\mu F_{\mu\nu} W^\nu
\,,
\end{align}
where $W_{\mu\nu}=2D_{[\mu} W_{\nu]}$ and $D_\mu = \partial_\mu - iQ A_\mu$. Note that the last term is a nonminimal interaction which is for now added with an arbitrary coefficient $q$. It is fixed to be $q=1$ by the Higgs mechanism. If we want to keep the transversality constraint in its simplest form we have to add a nonminimal interaction to \eqref{completeh} with $q=1$. Let us consider a more general action where a multiplet of massive spin-one fields $A_{BB'}$ can exhibit electromagnetic or Yang-Mills interactions (the trace over the gauge group indices is assumed)
\begin{align}\label{spinoneYM}
\mathcal{L} = -\tfrac{1}{2}D_{AA'}A_{BB'}D^{AA'}A^{BB'} + \tfrac{1}{2}D_{AA'}A_{BB'}D^{BB'}A^{AA'} - \tfrac{1}{2}m^2A_{AA'}A^{AA'} + \tfrac{q}{2}A^{AA'}F_{ABA'B'}A^{BB'}\, ,
\end{align}
the equation of motion being
\begin{align}
\square A_{AA'} - D_{BB'}D_{AA'}A^{BB'} - m^2 A_{AA'} + qF_{ABA'B'}A^{BB'} = 0 \, .
\label{eq:spin1_EM_EOM_(1,0)}
\end{align}
One can restore the transversality 
\begin{align}
D^{AA'}A_{AA'} = 0
\label{eq:spin1_EM_transverse_constraint}
\end{align}
for $q=1$ provided the vacuum equation of motion
\begin{align}
D^{BB'}F_{ABA'B'} = 0 \, ,
\end{align}
is satisfied for the gauge field. Indeed, the complete system \eqref{completeh} also described the dynamics of the gauge field. 

\subsubsection{Chiralization}
\label{sec:}

As the first step of chiralization, we define a new field
\begin{align}
\varphi_{AB} := m^{-1}D_{(A|A'|}A_{B)}^{\phantom{B)}A'} \, .
\label{eq:spin1_EM_def_(2,0)}
\end{align}
With the help of this definition, as we did previously, we can rewrite the equation of motion \eqref{eq:spin1_EM_EOM_(1,0)} as the first-order one
\begin{align}
2mD^{B}_{\phantom{B}A'}\varphi_{AB} - F_{AB}A^{B}_{\phantom{B}A'} - m^2A_{AA'} = 0 \, .
\end{align}
We can read this equation as the definition of $A_{AA'}$
\begin{align}
A_{AA'} = 2m^{-1}M_{A}^{\phantom{A}B}D^{C}_{\phantom{C}A'}\varphi_{BC} \, ,
\label{eq:spin1_EM_(1,1)_fct_(2,0)}
\end{align}
where we defined the matrix $M$ such that\footnote{It is always possible since the field strength $F_{AB}$ should be assumed small in the right units, otherwise our effective field theory should not even make sense since the fields whose magnitude is comparable to the mass of the object can destroy it or to lead to strong field effects where perturbation theory breaks down and quantum corrections have also to be taken into account. } (note that $\epsilon\fdu{A}{B}\equiv \delta_A^B$)
\begin{align}
\Big(\epsilon\fdu{A}{C} - \frac{1}{m^2}F_{A}^{\phantom{A}C}\Big)M_{C}^{\phantom{C}B} = \delta_{A}^{B} \, .
\label{eq:spin1_EM_def_M}
\end{align}
Plugging the expression for $A_{AA'}$ into the definition of $\varphi_{AB}$, \eqref{eq:spin1_EM_def_(2,0)}, we obtain the following second-order equation of motion
\begin{align}
m^2\varphi_{AB} = 2D_{(A|A'|}\Big(M_{B)}^{\phantom{B)}C}D^{DA'}\varphi_{CD}\Big) \, .
\label{eq:spin1_EM_chiral_EOM}
\end{align}
The equation can be obtained from a simple Lagrangian
\begin{align}
\mathcal{L} = D_{AA'}\varphi^{AB}M_{B}^{\phantom{B}C}D^{DA'}\varphi_{CD} + \tfrac{1}{2}m^2\varphi^{AB}\varphi_{AB} \, .
\end{align}
With some abuse of notation the Lagrangian is simply 
\begin{align}
    \mathcal{L} =\langle D\varphi|\frac{1}{1-F_+/m^2} |D\varphi\rangle +\tfrac12 m^2 \langle \varphi|\varphi\rangle \, ,
\end{align}
where $F_+\equiv F_{AB}$ is the selfdual component of the field strength. 

\paragraph{What happened to the transversality constraint?} Let us check that the transversality constraint in the case of electromagnetic interaction is trivial in the chiral formulation. We plug the expression for the $A_{BB'}$-field \eqref{eq:spin1_EM_(1,1)_fct_(2,0)}
into the transverse constraint \eqref{eq:spin1_EM_transverse_constraint} to find
\begin{align}
D^{AA'}\big(M_{A}^{\phantom{A}B}D^C_{\phantom{C}A'}\varphi_{BC}\big) = 0 \, .
\end{align}
It is not obvious that this condition is satisfied. Let us massage it. By using the definition of $M$ \eqref{eq:spin1_EM_def_M}, we can replace $M$ with
\begin{align}
M_{A}^{\phantom{A}B} = \delta_{A}^{B} + \frac{1}{m^2}F_{A}^{\phantom{A}C}M_{C}^{\phantom{C}B} \, ,
\end{align}
so that the expression becomes
\begin{align}
D^{AA'}D^C_{\phantom{C}A'}\varphi_{AC} + \frac{1}{m^2}D^{AA'}\big(F_{A}^{\phantom{A}D}M_{D}^{\phantom{D}B}D^C_{\phantom{C}A'}\varphi_{BC}\big) &= 0 \\
\Leftrightarrow \quad -\frac{1}{2}\epsilon^{AC}\square\varphi_{AC} - \frac{1}{2}F^{AC}\varphi_{AC} + \frac{1}{m^2}D^{AA'}\big(F_{A}^{\phantom{A}D}M_{D}^{\phantom{D}B}D^C_{\phantom{C}A'}\varphi_{BC}\big) &= 0 \, .
\end{align}
The first term is trivially zero and we can move $F$ in the last term outside of the derivative because we treat the case of a vacuum gauge field. The final result is
\begin{align}
F^{AB}\Big(m^2\varphi_{AB} - 2D_{AA'}\big(M_{B}^{\phantom{B}C}D^{DA'}\varphi_{CD}\big)\Big) = 0 \, ,
\end{align}
which is satisfied thanks to the equation of motion \eqref{eq:spin1_EM_chiral_EOM}. It means that the transverse constraint in the chiral formulation is automatically solved. In Appendix \ref{app:gyro} we slightly generalize this example by changing the gyromagnetic ratio. 

\section{Spin-three-half}
\label{sec:}

\subsection{Free fields}
\label{sec:}
Let us start with the $4d$ Rarita-Schwinger action already in the spinorial language. The vector-spinor $\psi^\mu$ can be decomposed into $(2,1)$, $(1,2)$, $(1,0)$ and $(0,1)$ representations of $sl(2,\mathbb{C})$, which corresponds to $\psi_{ABA'}$, its conjugate $\bar{\psi}_{AA'B'}$ and two auxiliary spinors: $\xi_{A}$ and its conjugated $\bar{\xi}_{A'}$. The Lagrangian reads
\begin{align}
\mathcal{L} &= \sqrt{2}\bar{\psi}^{AA'B'}\partial^{C}_{\phantom{C}A'}\psi_{ACB'} + \tfrac{1}{2}m\Big(\psi^{ABA'}\psi_{ABA'} - \bar{\psi}^{AA'B'}\bar{\psi}_{AA'B'}\Big) \notag \\
&\qquad - 3\sqrt{2}\bar{\xi}^{A'}\partial_{AA'}\xi^{A} + 3m\Big(\xi^{A}\xi_{A} - \bar{\xi}^{A'}\bar{\xi}_{A'}\Big) + \sqrt{2}\Big(\psi^{ABA'}\partial_{AA'}\xi_{B} + \bar{\psi}^{AA'B'}\partial_{AA'}\bar{\xi}_{B'}\Big) \, ,
\label{eq:spin3/2_free_Lagrangian}
\end{align}
where the coefficients are chosen in order to find the wanted constraints: the vanishing (suicide) of the auxiliary fields $\xi^A$, $\bar{\xi}^{A'}$ and the transverse constraint $\partial^{CC'}\psi_{ACC'} = 0$. The equations of motion obtained from this Lagrangian are
\besubeqs
\begin{align}
{E^{\psi}}_{ABA'} &:= m\psi_{ABA'} + \sqrt{2}\partial_{(A}^{\phantom{(A}B'}\bar{\psi}_{B)A'B'} + \sqrt{2}\partial_{(A|A'|}\xi_{B)} = 0 \, ,
\label{eq:spin3/2_free_EOM_1} \\[2mm]
{E^{\bar{\psi}}}_{AA'B'} &:= -m\bar{\psi}_{AA'B'} + \sqrt{2}\partial^{C}_{\phantom{C}(A'}\psi_{|AC|B')} + \sqrt{2}\partial_{A(A'}\bar{\xi}_{B')} = 0 \, ,
\label{eq:spin3/2_free_EOM_2} \\[2mm]
{E^{\xi}}_{A} &:= 6m\xi_{A} - 3\sqrt{2}\partial_{AA'}\bar{\xi}^{A'} - \sqrt{2}\partial^{CC'}\psi_{ACC'} = 0 \, ,
\label{eq:spin3/2_free_EOM_3} \\[2mm]
{E^{\bar{\xi}}}_{A'} &:= -6m\bar{\xi}_{A'} - 3\sqrt{2}\partial_{AA'}\xi^{A} - \sqrt{2}\partial^{CC'}\bar{\psi}_{CC'A'} = 0 \, .
\label{eq:spin3/2_free_EOM_4}
\end{align}
\esubeqs
The constraints can be obtained as combinations of the equations of motion and their derivatives. The expression
\begin{align}\label{constrflatthree}
\partial^{BB'}{E^{\psi}}_{ABB'} + \tfrac{\sqrt{2}}{2}m{E^{\xi}}_{A} + \tfrac{1}{2}\partial_{A}^{\phantom{A}A'}{E^{\bar{\xi}}}_{A'}&=3\sqrt{2}m^2 \xi_A
\end{align}
gives, when equations of motion are satisfied, the constraint $\xi_A = 0$. Equivalently, the following expression
\begin{align}\label{consxibar}
\partial^{BB'}{E^{\bar{\psi}}}_{BB'A'} - \tfrac{\sqrt{2}}{2}m{E^{\bar{\xi}}}_{A'} + \tfrac{1}{2}\partial^{A}_{\phantom{A}A'}{E^{\xi}}_{A}=-3\sqrt{2}m^2 \bar{\xi}_{A'}
\end{align}
gives on-shell the constraint $\bar{\xi}_{A'} = 0$. 
By using these constraints in the equations of motion, we obtain from \eqref{eq:spin3/2_free_EOM_1} and \eqref{eq:spin3/2_free_EOM_2} the two Dirac-like equations for the main fields
\besubeqs
\begin{align}
m\psi_{ABA'} + \sqrt{2}\partial_{(A}^{\phantom{(A}B'}\bar{\psi}_{B)A'B'} &= 0 \, , \\[2mm]
-m\bar{\psi}_{AA'B'} + \sqrt{2}\partial^{C}_{\phantom{C}(A'}\psi_{|AC|B')} &= 0 \, ,
\label{eq:spin3/2_free_Dirac_EOM_conj}
\end{align}
\esubeqs
and from \eqref{eq:spin3/2_free_EOM_3} and \eqref{eq:spin3/2_free_EOM_4} we get the transverse constraints
\besubeqs
\begin{align}
\partial^{CC'}\psi_{ACC'} &= 0 \, ,
\label{eq:spin3/2_free_transverse_constraint} \\
\partial^{CC'}\bar{\psi}_{CC'A'} &= 0 \, .
\label{eq:spin3/2_free_transverse_constraint_conj}
\end{align}
\esubeqs

\subsubsection{Chiralization}
\label{sec:}
Now, we will apply the procedure of chiralization in order to obtain the chiral description of the massive spin-$3/2$ field starting from the symmetric one. Let us begin by considering the equations of motion \eqref{eq:spin3/2_free_EOM_2} and \eqref{eq:spin3/2_free_EOM_3}, respectively, as a definition of $\bar{\psi}_{AA'B'}$ and $\xi_A$. Then, we use these definitions in the two other equations of motion \eqref{eq:spin3/2_free_EOM_1}, \eqref{eq:spin3/2_free_EOM_4} in order to obtain two equations of motion for $\psi_{ABA'}$ and $\bar{\xi}_{A'}$
\besubeqs
\begin{align}
m\psi_{ABA'} - m^{-1}\square\psi_{ABA'} + \tfrac{4}{3}m^{-1}\partial_{(A|A'|}\partial^{CC'}\psi_{B)CC'} &= 0 \, ,
\label{eq:spin3/2_free_EOM_(2,1)_2nd_order} \\
\bar{\xi}_{A'} &= 0 \, .
\end{align}
\esubeqs
The second equation is the ``suicide'' of the auxiliary field.\footnote{This should not be too surprising since solving for the $\xi_A$-field to plug it into another equation is equivalent to taking a linear combination of the equations from which $\xi_A$ disappears, which gives the same effect as \eqref{consxibar} here. } The first one is the second-order equation describing the main field $\psi_{ABA'}$. In order to obtain the chiral description, we define a new field
\begin{align}
\varphi_{ABC} := m^{-1}\partial_{(A}^{\phantom{(A}A'}\psi_{BC)A'} \, .
\label{eq:spin3/2_free_def_(3,0)}
\end{align}
The definition allows to rewrite the second-order equation of motion \eqref{eq:spin3/2_free_EOM_(2,1)_2nd_order} as the first-order one
\begin{align}
m\psi_{ABA'} + 2\partial^{C}_{\phantom{C}A'}\varphi_{ABC} = 0 \, .
\label{eq:spin3/2_free_EOM_(2,1)_(3,0)}
\end{align}
Finally, in order to obtain the chiral description, we swap the roles of the first-order equation \eqref{eq:spin3/2_free_EOM_(2,1)_(3,0)} and the definition \eqref{eq:spin3/2_free_def_(3,0)}. By using the definition of $\psi_{ABA'}$ \eqref{eq:spin3/2_free_EOM_(2,1)_(3,0)} in the first-order equation of motion \eqref{eq:spin3/2_free_def_(3,0)}, we obtain the following second-order equation of motion
\begin{align}
\big(\square - m^2\big)\varphi_{ABC} = 0 \, ,
\end{align}
which is well the Klein-Gordon equation describing a massive spin-$3/2$ field. The corresponding Lagrangian is simply
\begin{align}
\mathcal{L} = \tfrac{1}{2}\varphi^{ABC}\big(\square-m^2\big)\varphi_{ABC} \, .
\end{align}
Therefore, this result shows that the chiral and symmetric approaches to spin-$3/2$ are equivalent.

\paragraph{What happened to the transversality constraints?} 
The expression of the $(2,1)$-field $\psi_{ABA'}$ in terms of the $(3,0)$-field $\varphi_{ABC}$ \eqref{eq:spin3/2_free_EOM_(2,1)_(3,0)} can now be plugged into the transversality constraint \eqref{eq:spin3/2_free_transverse_constraint}, giving
\begin{align}
\epsilon^{AB}\square\varphi_{ABC} = 0 \, ,
\end{align}
which is trivially satisfied. We can also check the second transversality constraint. The equation \eqref{eq:spin3/2_free_Dirac_EOM_conj} can be rewritten as
\begin{align}
\bar{\psi}_{AA'B'} = \sqrt{2}m^{-1}\partial^{C}_{\phantom{C}(A'}\psi_{|AC|B')} \, .
\end{align}
By using it to replace $\bar{\psi}_{AA'B'}$ in the conjugate transverse constraint \eqref{eq:spin3/2_free_transverse_constraint_conj}
we obtain
\begin{align}
\epsilon^{AB}\square\psi_{ABA'} + 2\partial^{A}_{\phantom{B}A'}\partial^{BB'}\psi_{ABB'} = 0 \, .
\end{align}
The first term is trivially zero and the second one is zero by using the transversality constraint \eqref{eq:spin3/2_free_transverse_constraint}. 

\subsection{Einstein spaces}
\label{sec:}
For illustrative purposes we consider the simplest interaction of the massive spin three-half field, which turns out to be the gravitational one. It is also possible to perform the chiralization as to reveal the nonminimal interactions that restore parity. Let us begin by replacing $\pl$ with $\nabla$ in the Lagrangian \eqref{eq:spin3/2_free_Lagrangian}
\begin{align}
\mathcal{L} &= \sqrt{2}\bar{\psi}^{AA'B'}\nabla^{C}_{\phantom{C}A'}\psi_{ACB'} + \tfrac{1}{2}m\Big(\psi^{ABA'}\psi_{ABA'} - \bar{\psi}^{AA'B'}\bar{\psi}_{AA'B'}\Big) \notag \\
&\qquad - 3\sqrt{2}\bar{\xi}^{A'}\nabla_{\!\! AA'}\xi^{A} + 3m\Big(\xi^{A}\xi_{A} - \bar{\xi}^{A'}\bar{\xi}_{A'}\Big) + \sqrt{2}\Big(\psi^{ABA'}\nabla_{\!\! AA'}\xi_{B} + \bar{\psi}^{AA'B'}\nabla_{\!\! AA'}\bar{\xi}_{B'}\Big) \, .
\label{eq:spin3/2_free_Lagrangian_curved_ST}
\end{align}
We recall that the commutators of covariant derivatives were spelled out in section \ref{sec:half}. The equations of motion obtained from this Lagrangian are
\besubeqs
\begin{align}
{E^{\psi}}_{ABA'} &:= m\psi_{ABA'} + \sqrt{2}\nabla_{\!\! (A}^{\phantom{\!\! (A}B'}\bar{\psi}_{B)A'B'} + \sqrt{2}\nabla_{\!\! (A|A'|}\xi_{B)} = 0 \, ,
\label{eq:spin3/2_free_curved_ST_EOM_1} \\[2mm]
{E^{\bar{\psi}}}_{AA'B'} &:= -m\bar{\psi}_{AA'B'} + \sqrt{2}\nabla^{C}_{\phantom{C}(A'}\psi_{|AC|B')} + \sqrt{2}\nabla_{\!\! A(A'}\bar{\xi}_{B')} = 0 \, ,
\label{eq:spin3/2_free_curved_ST_EOM_2} \\[2mm]
{E^{\xi}}_{A} &:= 6m\xi_{A} - 3\sqrt{2}\nabla_{\!\! AA'}\bar{\xi}^{A'} - \sqrt{2}\nabla^{CC'}\psi_{ACC'} = 0 \, ,
\label{eq:spin3/2_free_curved_ST_EOM_3} \\[2mm]
{E^{\bar{\xi}}}_{A'} &:= -6m\bar{\xi}_{A'} - 3\sqrt{2}\nabla_{\!\! AA'}\xi^{A} - \sqrt{2}\nabla^{CC'}\bar{\psi}_{CC'A'} = 0 \, .
\label{eq:spin3/2_free_curved_ST_EOM_4}
\end{align}
\esubeqs
The form of the constraint is inherited from the flat space \eqref{constrflatthree}
\begin{align}
\nabla^{BB'}{E^{\psi}}_{ABB'} + \tfrac{\sqrt{2}}{2}m{E^{\xi}}_{A} + \tfrac{1}{2}\nabla_{\!\! A}^{\phantom{\!\! A}A'}{E^{\bar{\xi}}}_{A'} = 0 \, ,
\label{eq:free_curved_ST_constraint}
\end{align}
and gives
\begin{align}
3\sqrt{2}\big(m^2+R\big)\xi_A - \tfrac{\sqrt{2}}{2}R_{ABA'B'}\bar{\psi}^{BA'B'} = 0 \, ,
\label{eq:spin3/2_free_curved_ST_constraint}
\end{align}
which does not imply the vanishing of the auxiliary field per se. However, for the vacuum Einstein spacetimes we have
\begin{align}
R_{ABA'B'} &= 0\,,  &
R &= \Lambda\,,
\end{align}
where $\Lambda$ is related to the cosmological constant. The constraint becomes
\begin{align}
3\sqrt{2}\big(m^2+\Lambda\big)\xi_A = 0 \, ,
\end{align}
from which we deduce the expected constraint $\xi_A = 0$ only if
\begin{align}
m^2 \ne -\Lambda \, .
\end{align}
The latter is, obviously, the usual massless point where the Rarita-Schwinger action becomes gauge invariant, idem for $\bar{\xi}^{A'}$ and $\bar{\psi}^{ABA'}$. Finally, applying this constraint in the third and fourth equations of motion gives the transverse constraint for the main field $\nabla^{CC'}\psi_{ACC'}=0$. Therefore, the minimal coupling of the massive spin-$3/2$ to an Einstein spacetime is sufficient to have the right number of degrees of freedom.

\subsubsection{Chiralization}
\label{sec:}

Now, let us apply the procedure of chiralization in the case of Einstein spacetimes. By using the equations \eqref{eq:spin3/2_free_curved_ST_EOM_2} and \eqref{eq:spin3/2_free_curved_ST_EOM_3} as the definitions of, respectively, $\bar{\psi}_{AA'B'}$ and $\xi_A$ and plugging them into the equations of motion \eqref{eq:spin3/2_free_curved_ST_EOM_1} and \eqref{eq:spin3/2_free_curved_ST_EOM_4}, we obtain
\begin{align}
m^2\psi_{ABA'} - \square\psi_{ABA'} + \tfrac{4}{3}\nabla_{\!\!(A|A'|}\nabla^{CC'}\psi_{B)CC'} + 4\Lambda\psi_{ABA'} + C_{ABCD}\psi^{CD}_{\phantom{CD}A'} &= 0 \,, \label{eq:spin3/2_gravit_EOM_second_(2,1)} \\
\big(m^2 + \Lambda\big)\bar{\xi}_{A'} &= 0 \, .
\end{align}
The second one is the ``suicide'' of the auxiliary field only if
$m^2 \ne -\Lambda$. Next, we define a completely chiral field
\begin{align}
\varphi_{ABC} := m^{-1}\nabla_{\!\!(A}^{\phantom{\!\!(A}A'}\psi_{BC)A'} \, .
\label{eq:spin3/2_gravit_def_(3,0)}
\end{align}
We can rewrite the second-order equation \eqref{eq:spin3/2_gravit_EOM_second_(2,1)} as the following first-order one
\begin{align}
m\psi_{ABA'} + 2\nabla^{C}_{\phantom{C}A'}\varphi_{ABC} + m^{-1}\Lambda\psi_{ABA'} + m^{-1}C_{ABCD}\psi^{CD}_{\phantom{CD}A'} = 0 \, .
\end{align}
As usual, we swap the definition of an auxiliary field and the dynamical equation in order to express $\psi_{ABA'}$ as
\begin{align}
\psi_{ABA'} = -2m^{-1}M_{AB}^{\phantom{AB}CD}\nabla^{E}_{\phantom{E}A'}\varphi_{CDE} \, ,
\label{eq:spin3/2_gravit_def_(2,1)}
\end{align}
where $M_{ABCD}$ is defined such that
\begin{align}
\bigg(\Big(1+\frac{\Lambda}{m^2}\Big)\epsilon_{(A}^{\phantom{(A}E}\epsilon_{B)}^{\phantom{B)}F} + \frac{1}{m^2}C_{AB}^{\phantom{AB}EF}\bigg)M_{EF}^{\phantom{EF}CD} = \delta_{(A}^{C}\delta_{B)}^{D} \, .
\label{eq:spin3/2_grav_def_M}
\end{align}
Note that $M_{ABCD}$ does not have the same symmetries as $C_{ABCD}$. By plugging \eqref{eq:spin3/2_gravit_def_(2,1)} into \eqref{eq:spin3/2_gravit_def_(3,0)} we obtain
\begin{align}
m^2\varphi_{ABC} + 2\nabla_{\!\!(A}^{\phantom{\!\!(A}A'}\Big(M^{DF}_{\phantom{DF}BC)}\nabla^{E}_{\phantom{E}A'}\varphi_{DEF}\Big) = 0 \, ,
\label{eq:spin3/2_grav_EOM_(3,0)}
\end{align}
which comes from a simple Lagrangian
\begin{align}
\mathcal{L} = \tfrac{1}{2}m^2\varphi^{ABC}\varphi_{ABC} - \nabla^{AA'}\varphi_{ABC}M^{BC}_{\phantom{BC}DF}\nabla_{\!\!EA'}\varphi^{DEF} \, .
\end{align}
It is the Lagrangian density describing the free massive spin-three-half field in an Einstein spacetime in the chiral approach for the case $m^2 \ne -\Lambda$. Again, with some abuse of notation the Lagrangian is simply 
\begin{align}
    \mathcal{L} =\langle \nabla\varphi|\frac{1}{1+\Lambda/m^2+C_+/m^2} |\nabla\varphi\rangle +\tfrac12 m^2 \langle \varphi|\varphi\rangle\,,
\end{align}
where $C_+\equiv C_{ABCD}$ is the selfdual component of the Weyl tensor.

\paragraph{What happened to the transversality constraints?} 
Let us check that the constraints are preserved during the chiralization procedure. We can set the auxiliary fields to zero in the equations to get
\besubeqs
\begin{align}
m\psi_{ABA'} + \sqrt{2}\nabla_{\!(A}^{\phantom{\!(A}B'}\bar{\psi}_{B)A'B'} &= 0\,,
\label{eq:spin3/2_grav_EOM_1_with_constr} \\[2mm]
\bar{\psi}_{AA'B'} &= \sqrt{2}m^{-1}\nabla^{C}_{\phantom{C}(A'}\psi_{|AC|B')}\,,
\label{eq:spin3/2_inter_EOM_2_with_constr} \\[2mm]
\nabla^{CC'}\psi_{ACC'} &= 0\,,
\label{eq:spin3/2_grav_transverse_constraint} \\[2mm]
\nabla^{CC'}\bar{\psi}_{CC'A'} &= 0 \, ,
\label{eq:spin3/2_grav_transverse_constraint_conj}
\end{align}
\esubeqs
where the two first equations are the Dirac equations for the main field and its conjugate, and the two last ones are the transversality constraints. The first step of the chiralization consists in using the second Dirac equation to replace $\bar{\psi}$ with $\psi$ in the rest of the equations. By doing this in the first Dirac equation, it gives the second-order equation \eqref{eq:spin3/2_gravit_EOM_second_(2,1)}. Let us see how the conjugate transverse constraint \eqref{eq:spin3/2_grav_transverse_constraint_conj} looks like after this replacement
\begin{align}
\nabla^{CC'}\nabla^{D}_{\phantom{C}A'}\psi_{CDC'} + \nabla^{CC'}\nabla^{D}_{\phantom{C}C'}\psi_{CDA'} = 0 \, ,
\end{align}
which gives, by using the Fierz identity on the first term,
\begin{align}
-2\nabla^{C}_{\phantom{C}C'}\nabla^{DC'}\psi_{CDA'} + \nabla^{D}_{\phantom{D}A'}\nabla^{CC'}\psi_{CDC'} = 0 \, .
\end{align}
The second term is zero because of the transverse constraint \eqref{eq:spin3/2_grav_transverse_constraint}. By developing the first one we obtain
\begin{align}
R_{ABA'B'}\psi^{ABB'} = 0 \, ,
\end{align}
which is trivially satisfied in an Einstein spacetime. The transverse constraint of the conjugate field expressed in terms of the ``main'' $\psi_{ABA'}$-field is trivial, as expected to keep the right number of degrees of freedom.

Let us check that the transverse constraint \eqref{eq:spin3/2_grav_transverse_constraint} is trivial in the final step of chiralization. The expression of the $\psi_{ABA'}$-field in terms of the chiral $\varphi_{ABC}$-field is given by \eqref{eq:spin3/2_gravit_def_(2,1)}. 
By using this expression to replace the $\psi_{ABA'}$-field in the transverse constraint \eqref{eq:spin3/2_grav_transverse_constraint}, we obtain
\begin{align}
\nabla^{BB'}\big(M_{AB}^{\phantom{AB}CD}\nabla^{E}_{\phantom{E}B'}\varphi_{CDE}\big) = 0 \, .
\end{align}
It is not obvious that this constraint is satisfied. By using the definition of $M$ \eqref{eq:spin3/2_grav_def_M}, we can develop the constraint as 
\begin{align}
m^2\nabla^{BB'}\nabla^{C}_{\phantom{C}B'}\varphi_{ABC} - \nabla^{BB'}\big(C_{ABFG}M^{FGCD}\nabla^{E}_{\phantom{E}B'}\varphi_{CDE}\big) = 0 \, .
\end{align}
In the second term we can use the Bianchi identity $\nabla^{AA'}C_{ABCD} = 0$ to move the $C$ outside of the derivative. We obtain
\begin{align}
C_{A}^{\phantom{A}BCD}\Big(m^2\varphi_{BCD} + 2\nabla_{\!B}^{\phantom{\!B}B'}\big(M_{CD}^{\phantom{CD}EF}\nabla^{G}_{\phantom{G}B'}\varphi_{EFG}\big)\Big) = 0 \, ,
\end{align}
which is trivially satisfied because of the chiral equation of motion \eqref{eq:spin3/2_grav_EOM_(3,0)}. It means that the transverse constraint expressed in terms of the chiral field is automatically satisfied.

\section{Spin-two}
\label{sec:}
For the case of a massive spin-two field the relation between symmetric and chiral formulations is less than obvious.  

\subsection{Free fields in the symmetric approach}
\label{sec:}

The spin-two case is well-known since Fierz and Pauli \cite{Fierz:1939ix}, who discovered that the massive spin-two equations of motion can be obtained from a Lagrangian with one auxiliary scalar field. Let us reproduce this result for completeness. The correct Lagrangian reads 
\begin{equation}
\mathcal{L} = -\tfrac{1}{2}\partial_{\alpha}h_{\mu\nu}\partial^{\alpha}h^{\mu\nu} + \partial_{\alpha}h^{\alpha\mu}\partial^{\beta}h_{\beta\mu} - \tfrac{1}{2}m^2h_{\mu\nu}h^{\mu\nu} - \partial_{\alpha}h^{\alpha\beta}\partial_{\beta}\xi + \tfrac{3}{4}\partial^{\alpha}\xi\partial_{\alpha}\xi + \tfrac{3}{2}m^2\xi^2 \text{ ,}
\label{eq:spin2_action_Lorentz}
\end{equation}
where the relative coefficients are chosen to guarantee the constraints that will give the correct number of degrees of freedom for a massive spin-two field. We will use this action as the starting point to transfer the theory to the chiral formulation. The equations of motion are
\begin{equation}
E_{\mu\nu}^h := -m^2h_{\mu\nu} + \square h_{\mu\nu} - \partial_{\mu}\partial^{\alpha}h_{\alpha\nu} - \partial_{\nu}\partial^{\alpha}h_{\alpha\mu} + \partial_{\mu}\partial_{\nu}\xi + \tfrac{1}{2}\eta_{\mu\nu}\partial_{\alpha}\partial_{\beta}h^{\alpha\beta} - \tfrac{1}{4}\eta_{\mu\nu}\square\xi = 0 \, ,
\label{eq:EOM_h_Lorentz}
\end{equation}
and
\begin{equation}
E^{\xi} := 3m^2\xi + \partial_{\alpha}\partial_{\beta}h^{\alpha\beta} - \tfrac{3}{2}\square\xi = 0 \, .
\label{eq:EOM_chi_Lorentz}
\end{equation}
In order to get the wanted on-shell conditions one needs to play with the Lagrangian equations. First, we get the following low-derivative consequence
\begin{equation}\label{semitransverse}
\partial^{\nu}E_{\mu\nu}^h + \frac{1}{2} \partial_{\mu}E^{\xi} = -m^2 (\partial^{\nu}h_{\mu\nu} - \tfrac32\partial_{\mu}\xi )\, .
\end{equation}
One more low-derivative consequence is found via
\begin{equation}
m^2E^{\xi} + \frac{1}{2}\partial^{\mu}\partial_{\mu}E^{\xi} + \partial^{\mu}\partial^{\nu}E^h_{\mu\nu} = 3m^4\xi \text{ .}
\end{equation}
Therefore, on the equations of motion we have 
the suicide $\xi=0$, which, upon plugging into \eqref{semitransverse} yields the transversality constraint $\partial^{\nu}h_{\mu\nu} = 0$. When we use these two in the initial operator $E^h_{\mu\nu}$ we obtain the Klein-Gordon equation
\begin{equation}
\big(\square - m^2\big)h_{\mu\nu} = 0 \,.
\end{equation}
Therefore, $h_{\mu\nu}$ carries $10-1-4=5$ degrees of freedom as the massive $s=2$ field should.

In order to make a transfer to the chiral approach we need first to rewrite the action in the spinorial language. A traceless symmetric field $h_{\mu\nu}$ becomes a $(2,2)$ spin-tensor $h_{AB,A'B'}$ of $sl(2,\mathbb{C})$. Nothing dramatic happens to the scalar field $\xi$. The Lagrangian \eqref{eq:spin2_action_Lorentz} can be rewritten as
\begin{align}
\mathcal{L} = -\tfrac{1}{2}\partial_{CC'}h_{ABA'B'}\partial^{CC'}h^{ABA'B'} &+ \partial_{CC'}h^{ACA'C'}\partial^{BB'}h_{ABA'B'} - \tfrac{1}{2}m^2h_{ABA'B'}h^{ABA'B'} \notag \\
 &- \partial_{CC'}h^{BCB'C'}\partial_{BB'}\xi + \tfrac{3}{4}\partial^{CC'}\xi\partial_{CC'}\xi + \tfrac{3}{2}m^2\xi^2 \, .
\label{eq:spin2_action_spinor}
\end{align}
Indeed, the equations of motion we obtain are
\besubeqs
\begin{align}
-m^2h_{ABA'B'} + \square h_{ABA'B'} + \partial_{(A|A'}\partial_{|B)B'}\xi -2\partial_{(A|{\color{red}(}A'{\color{red}|}}\partial^{DD'}h_{|B)D{\color{red}|}B'{\color{red})}D'} &= 0 \,, \label{eq:EOM_h_spinor} \\[2mm]
3m^2\xi - \tfrac{3}{2}\square\xi + \partial^{AA'}\partial^{CC'}h_{ACA'C'} &= 0 \, .
\label{eq:EOM_chi_spinor}
\end{align}
\esubeqs
These equations are the same than \eqref{eq:EOM_h_Lorentz} and \eqref{eq:EOM_chi_Lorentz}, remembering that the symmetry over spinor indices corresponds to tracelessness in vector language.

\subsubsection{Chiralization}
\label{sec:}

As different from the spin-one example, the transfer to the chiral formulation cannot be accomplished in one step. We will have to perform it in two steps. The main dynamical field is of type $(2,2)$. We can relate it to a type-$(3,1)$ field via one derivative and the latter one can be related to the wanted $(4,0)$-field via one derivative. We begin by defining the following auxiliary fields of the intermediate formulation
\besubeqs
\begin{align}
\phi_{ABCA'} &:= m^{-1}\partial_{(C}^{\phantom{(C}B'}h_{AB)A'B'} \,, \label{eq:spin2_def_(3,1)} \\
\psi_{AA'} &:= m^{-1}\partial^{BB'}h_{ABA'B'} - \tfrac{3}{2}m^{-1}\partial_{AA'}\xi \text{ .} \label{eq:spin2_def_(1,1)}
\end{align}
\esubeqs
Using these definitions, we can rewrite the equations \eqref{eq:EOM_h_spinor} and \eqref{eq:EOM_chi_spinor} respectively as the following first-order equations
\besubeqs
\begin{align}
h_{ABA'B'} &= -2m^{-1}\partial^C_{\phantom{C}(A'|}\phi_{ABC|B')} - \tfrac{2}{3}m^{-1}\partial_{(A|{\color{red}(}A'{\color{red}|}}\psi_{|B){\color{red}|}B'{\color{red})}} \,, \label{eq:spin2_EOM_h_phi_psi} \\
\xi &= -\tfrac{1}{3}m^{-1}\partial^{AA'}\psi_{AA'} \text{ .}
\label{eq:spin2_EOM_chi_psi}
\end{align}
\esubeqs
Now, we can take these first-order equations as definitions of the original fields and rewrite \eqref{eq:spin2_def_(3,1)} and \eqref{eq:spin2_def_(1,1)}, respectively, as the second-order equations
\besubeqs
\begin{align}
{E^{\phi}}_{ABCA'} := 3m^2\phi_{ABCA'} - 3\square\phi_{ABCA'} + 3\partial_{(A|A'}\partial^{DD'}\phi_{|BC)DD'} + \partial_{(A|A'|}\partial_{B}^{\phantom{B}B'}\psi_{C)B'} &= 0 \,, \label{eq:EOM_phi} \\
{E^{\psi}}_{AA'} :=3m^2\psi_{AA'} + 3\partial^B_{\phantom{B}A'}\partial^{CC'}\phi_{ABCC'} + \square\psi_{AA'} - \partial_{AA'}\partial^{BB'}\psi_{BB'} &= 0 \text{ .} \label{eq:EOM_psi}
\end{align}
\esubeqs
These equations are equations of motion for $\phi_{ABCC'}$ and $\psi_{AA'}$. As a result we shifted the field content from $(2,2)\oplus (0,0)$ to a more chiral $(3,1)\oplus(1,1)$. We will now try again the same manipulation to obtain completely chiral fields. Let us define the following fields
\besubeqs
\begin{align}
\varphi_{ABCD} &:= m^{-1}\partial_{(A}^{\phantom{(A}B'}\phi_{BCD)B'}\,,\label{eq:spin2_def_(4,0)} \\
\Psi_{AB} &:= m^{-1}\partial^{DD'}\phi_{ABDD'} - \tfrac{2}{3}m^{-1}\partial_{(A}^{\phantom{(A}B'}\psi_{B)B'} \,.\label{eq:spin2_def_(2,0)}
\end{align}
\esubeqs
Using these definitions, we can rewrite the equations \eqref{eq:EOM_phi} and \eqref{eq:EOM_psi}, respectively, as the first-order equations
\besubeqs
\begin{align}
\phi_{ABCA'} &= -2m^{-1}\partial^D_{\phantom{D}A'}\varphi_{ABCD} + \tfrac{1}{2}m^{-1}\partial_{(A|A'}\Psi_{|BC)} \,, \label{eq:spin2_EOM_phi_varphi_Psi} \\
\psi_{AA'} &= -m^{-1}\partial^B_{\phantom{B}A'}\Psi_{AB} \, . \label{eq:spin2_EOM_psi_Psi}
\end{align}
\esubeqs
Now, we treat these first-order equations as definitions of the stage-two fields and rewrite the first-order relations \eqref{eq:spin2_def_(4,0)} and \eqref{eq:spin2_def_(2,0)} respectively as the following second-order equations
\begin{align}
\big(\square - m^2\big)\varphi_{ABCD} &= 0 \,,
\label{eq:spin2_EOM_varphi} \\
\Psi_{AB} &= 0 \,. \label{eq:spin2_EOM_Psi}
\end{align}
It should not escape one's notice that the second auxiliary field, of type $(2,0)$, eliminates itself from the system, so-called ``suicide'' of an auxiliary field. The first equation is just the Klein-Gordon equation describing the propagation of the massive field $\varphi_{ABCD}$ on the Minkowski spacetime. The field $\varphi_{ABCD}$, being totally symmetric, contains exactly the five physical components that a massive spin-two particle needs. The action is self-evident 
\begin{equation}
S = \tfrac{1}{2}\int\! \varphi^{ABCD}\big(\square - m^2\big)\varphi_{ABCD}\;\mathrm{d}^4x \,.\label{eq:action_spin2_spinor}
\end{equation}
The steps we have performed at the level of equations of motion can mutatis mutandis be implemented at the level of actions. 

\paragraph{What happened to the transversality constraints?} 

Let us analyze what happens to the constraints during the procedure of chiralization. In the case of spin $2$, there are two complete steps in the chiralization procedure, so we need to check the constraints at each step. At the beginning, the equations of motion are \eqref{eq:EOM_h_spinor} and \eqref{eq:EOM_chi_spinor}, from which we can extract the constraints $\xi=0$ and $\partial^{CC'}h_{ACA'C'} = 0$, 
as we shown previously. Then, we defined two new fields 
\eqref{eq:spin2_def_(3,1)}, \eqref{eq:spin2_def_(1,1)} of type $(3,1)$ and $(1,1)$, respectively, and we can write the $(2,2)$- and $(0,0)$-fields in terms of these new fields \eqref{eq:spin2_EOM_h_phi_psi}, \eqref{eq:spin2_EOM_chi_psi}. The new $(3,1)$- and $(1,1)$-fields satisfy the equations of motion \eqref{eq:EOM_phi} and \eqref{eq:EOM_psi}, 
from which we can deduce constraints. Indeed, the following combination of equations of motion gives
\begin{align}
m^2{E^{\psi}}_{AA'} + \tfrac{1}{3}\partial_{AA'}\partial^{BB'}{E^{\psi}}_{BB'} - \tfrac{1}{3}\square{E^{\psi}}_{AA'} - \partial^{C}_{\phantom{C}A'}\partial^{BB'}{E^{\phi}}_{ABCB'} = 3m^4\psi_{AA'} \, ,
\end{align}
which implies on-shell the suicide 
\begin{align}
\psi_{AA'} = 0 \, .
\label{eq:spin2_free_(1,1)=0}
\end{align}
It is the vanishing of the new auxiliary field. By using this constraint, we also find
\begin{align}
\partial^{CC'}{E^{\phi}}_{ABCC'} - \tfrac{2}{3}\partial_{(A}{}^{A'}{E^{\psi}}_{B)A'} = 3m^2\partial^{CC'}\phi_{ABCC'} \, ,
\end{align}
which is the transversality constraint for the $(3,1)$-field
\begin{align}
\partial^{CC'}\phi_{ABCC'} = 0 \, .
\label{eq:spin2_free_transverse_constraint_(3,1)}
\end{align}
Now, we know all the hidden (low derivative consequences) in this intermediate formulation. Let us check what the constraints on the old fields imply in terms of the new fields by using \eqref{eq:spin2_EOM_h_phi_psi} and \eqref{eq:spin2_EOM_chi_psi}. They become, respectively,
\besubeqs
\begin{align}
\xi = 0 \quad &\Leftrightarrow \quad \partial^{AA'}\psi_{AA'} = 0 \, , \\
\partial^{BB'}h_{ABA'B'} = 0 \quad &\Leftrightarrow \quad -\partial^{B}_{\phantom{B}A'}\partial^{CC'}\phi_{ABCC'} + \tfrac{1}{6}\partial_{AA'}\partial^{BB'}\psi_{BB'} + \tfrac{1}{3}\square\psi_{AA'} = 0 \, .
\end{align}
\esubeqs
The first expression is trivial because of the vanishing of the auxiliary $(1,1)$-field \eqref{eq:spin2_free_(1,1)=0}. The first term of the second expression vanishes because of the transverse constraint on the main $(3,1)$-field \eqref{eq:spin2_free_transverse_constraint_(3,1)}, and the two last terms are trivial because of the vanishing of the auxiliary $(1,1)$-field too. Therefore, it means that the constraints on the old fields in terms of the new ones do not imply more than what can be deduced from equations of motion for the new fields, as expected.

Now we check the relations between the intermediate and the chiral formulations. We need to check what the constraints for the $(3,1)$- and $(1,1)$-fields imply when they are expressed in terms of the new fields of type $(4,0)$ and $(2,0)$. These new fields are defined by \eqref{eq:spin2_def_(4,0)} and \eqref{eq:spin2_def_(2,0)}. The expression of the old $(3,1)$- and $(1,1)$-fields in terms of the new ones are given by \eqref{eq:spin2_EOM_phi_varphi_Psi} and \eqref{eq:spin2_EOM_psi_Psi}. 
The new fields satisfy the equations of motion \eqref{eq:spin2_EOM_varphi} and \eqref{eq:spin2_EOM_Psi}. 
The transverse constraint for the $(3,1)$-field \eqref{eq:spin2_free_transverse_constraint_(3,1)} expressed in terms of the new fields says
\begin{align}
3\epsilon^{CD}\square\varphi_{ABCD} + \square\Psi_{AB} = 0 \, .
\end{align}
The first term is trivially zero and the second one is zero because of the equation of motion, ``suicide'' of the auxiliary $(2,0)$-field \eqref{eq:spin2_EOM_Psi}. The vanishing of the auxiliary $(1,1)$-field \eqref{eq:spin2_free_(1,1)=0} expressed in terms of the new fields says
\begin{align}
\partial^{B}_{\phantom{B}A'}\Psi_{AB} = 0 \, ,
\end{align}
which is trivially satisfied because of the equations of motion again. Therefore, the constraints expressed in the chiral language are automatically satisfied.

From the Hamiltonian point of view the Fierz-Pauli system has $3+3$ vector-like constraints and $4$ scalar constraints in the $3+1$ language. All constraints are second class. Therefore, $10+10-(3+3+4)=10$ gives $5$ degrees of freedom. Covariantly this relies on the suicide $\xi=0$ and $\pl^\mu h_{\mu\nu}=0$. In the intermediate formulation the hidden low derivative consequences are $\pl^{MM'} \phi_{ABMM'}=0$ and $\psi_{AA'}=0$. The latter leave us with $5=8+4-4-3$ degrees of freedom again. It is clear that the action for these fields will lead to the second class constraints supporting this counting. Finally, in the last step before the chiral formulation $\Psi_{AB}=0$ is one of the equations (not a low derivative consequence) and $\varphi_{ABCD}$ does not need any constraints. Given that we can set $\psi_{AA'}=0$ from the very beginning, the intermediate formulation is in terms of the $(3,1)$-field that has $3$ transversality constraints. This way, we see the following chain of fields/constraints $5=10-4-1$, $5=8-3$ and $5=5-0$ (less and less fields and constraints as we move towards the chiral formulation).

\section{Conclusions and Discussion}
\label{sec:}
The main result of the paper is to show that the standard symmetric tensor formulation of massive fields with spin is equivalent to the recently proposed chiral formulation. This was first achieved at the free level and some examples of interactions have also been given. The main advantage of the chiral formulation is that it does not require an intricate (and also growing with spin) set of auxiliary fields. A general feature of the chiral formulation is that certain nonminimal interactions need to be present if the theory is parity invariant.

From the jet space point of view, the chiral fields provide a different, but equivalent to the standard one, parameterization of the solution space. From the Hamiltonian point of view the passage to the chiral variables allows one to (implicitly) solve the second class constraints without breaking Lorentz invariance. It would be interesting to apply the chiral approach within the presymplectic formalism, see e.g. \cite{Grigoriev:2016wmk, Grigoriev:2021wgw}, as to resolve some puzzles concerning the existence of the intrinsic Lagrangian formulations.   

Even though we worked mostly with the equations of motion, it is clear that there exist parent actions that relate the two formulations. It is also obvious that there should exist the invertible change of variables and the parent action for massive fields of arbitrary spin $s>2$, which are not covered in the text. However, the chain of auxiliary fields relating the symmetric tensors to the chiral fields grows with spin. It would be interesting to work it out in the future. It would also be interesting to find the chiral Lagrangians for the known consistent theories of massive spin-two fields, i.e. for massive (bi)gravity.

It should also be noted that various chiral, spinor-helicity, twistor etc. approaches quite often resort first to a complexification of the problem (e.g. the chiral fields are clearly complex in the Minkowski signature) where the actual real problem is to be selected by appropriate reality conditions. Therefore, certain aspects, e.g. the reality of the action, the boundness of the energy, etc., cannot be considered before the reality conditions are given, which is already nontrivial in the chiral formulation of gravity, see e.g. \cite{Delfino:2012zy}. Nevertheless, if the main goal is to compute amplitudes, the usage of some complexified form is not only possible but advisable. 

One of the main applications of massive higher spin fields is to look for effective theories that couple them to electromagnetic and gravitational interactions in the most simple and parity invariant way. Whenever the effective field theory is to describe the dynamics of black holes the notion of ``simple interaction'' is well-understood at least at the cubic level \cite{Arkani-Hamed:2017jhn, Guevara:2018wpp,Chung:2018kqs,Skvortsov:2023jbn}: the one that has the best high energy behavior also agrees with the direct predictions of general relativity. The main challenge is to find the complete effective field theory that completes this cubic interaction, see \cite{Cangemi:2022bew,Cangemi:2023bpe} for the progress at the quartic order.

\section*{Acknowledgments}
\label{sec:Aknowledgements}
The work of W.D. and E.S. was partially supported by the European Research Council (ERC) under the European Union’s Horizon 2020 research and innovation programme (grant agreement No 101002551). The work of W.D. was also supported by UMONS stipend ``Bourse d'encouragement doctorale FRIA/FRESH''. E.S. is grateful to Yuri Zinoviev for opening up the pandora box of massive higher spin fields and is also grateful to Maxim Grigoriev and Alexey Sharapov for numerous discussions on the topic.

\appendix

\section{First-order formulation of Proca theory}
\label{app:Proca}
It is instructive to compare the Chalmers-Siegel trick to the ordinary first-order formulation of Proca theory, which is applicable in any spacetime dimension $d$. The first-order action reads 
\begin{align}
    S[A,\Psi]&=\int -\tfrac12 F_{\mu\nu} \Psi^{\mu\nu} + \tfrac14 \Psi_{\mu\nu}\Psi^{\mu\nu} - \tfrac12 m^2 A^{\mu}A_{\mu} \, .
\end{align}
We can also eliminate $A_\mu$ via its equation of motion $\pl_\nu \Psi^{\mu\nu} - m^2 A^\mu=0$ to arrive at
\begin{align}
    S[\Psi]&=\int (\pl_\nu \Psi^{\mu\nu})(\pl_\lambda \Psi\fdu{\mu}{\lambda})+\tfrac12 m^2 \Psi_{\mu\nu}\Psi^{\mu\nu} \, ,
\end{align}
after a rescaling of $\Psi_{\mu\nu}$. The equation of motion 
\begin{align}
    E^{\mu\nu}&=\pl^\mu \pl_\lambda \Psi^{\nu\lambda}-\pl^\nu \pl_\lambda \Psi^{\mu\lambda}+m^2 \Psi^{\mu\nu}=0
\end{align}
implies $\pl_\nu E^{\mu\nu}=(\square - m^2) \pl_\lambda \Psi^{\mu\lambda}=0$. It is the divergence $A^\mu= \pl_\lambda \Psi^{\mu\lambda}$ that plays the role of the transverse vector field. In some sense $\pl_\lambda \Psi^{\mu\lambda}$ is just a way to represent any transverse vector in $d$-dimension without having to solve the constraint $\pl_\mu A^\mu=0$ in a non-local way as $A_\mu- \tfrac{1}{\square}\pl_\mu  \pl^\nu A_\nu$. The transverse components of $\Psi^{\mu\nu}$ do not propagate.  

To carry on the Hamiltonian analysis let us denote $C^i=\Psi^{0i}$, $B^{ij}=\Psi^{ij}$. The Lagrangian can be rewritten as
\begin{align}
    L&=+\tfrac 12(\dot{C}^i - \pl_k B^{ki})^2 -\tfrac12 (\pl_i C^i)^2-\tfrac{m^2}2 (C^i)^2+\tfrac{m^2}2 (B^{ij})^2  \, .
\end{align}
The primary constraint is $\Phi_1^{ij}=p^{ij}$. The Hamiltonian reads
\begin{align}
    H&= \tfrac12 (p_i)^2 +p_i \pl_k B^{ki} +\tfrac12 (\pl_i C^i)^2+\tfrac{m^2}2 (C^i)^2-\tfrac{m^2}2 (B^{ij})^2  \, .
\end{align}
The secondary constraint is $\Phi_2^{ij}=m^2 B^{ij}-\pl^{[i} p^{j]}$.
The Hamiltonian in the physical variables is
\begin{align}
    H&= \tfrac12 (p_i)^2 - \tfrac{1}{2m^2}(\pl_{[k} p_{i]})^2 +\tfrac12 (\pl_i C^i)^2+\tfrac{m^2}2 (C^i)^2 \, .
\end{align}
However, there is no obvious covariant action that leads to this Hamiltonian. 
The very first step of the analysis, the fact that there is a primary constraint, tells us that the action obtained from the first-order one is no better than the Proca action in that we still have constraints to deal with. On the other hand, the Chalmers-Siegel first-order action allows one to eliminate the second class constraints in a covariant way.

\section{Generic gyromagnetic ratio}
\label{app:gyro}
Let us unlock the gyromagnetic ratio by putting a free parameter in front of the nonminimal coupling in \eqref{spinoneYM}
\begin{align}
\mathcal{L} = -\tfrac{1}{2}D_{AA'}A_{BB'}D^{AA'}A^{BB'} + \tfrac{1}{2}D_{AA'}A_{BB'}D^{BB'}A^{AA'} + \frac{q}{2}A^{AA'}F_{ABA'B'}A^{BB'} - \tfrac{1}{2}m^2A_{AA'}A^{AA'} \, .
\label{eq:spin1_EMa_Lagrangian}
\end{align}
The equation of motion is
\begin{align}
E_{AA'} := \square A_{AA'} - D_{BB'}D_{AA'}A^{BB'} + \frac{q}{2}\big(F_{A}^{\phantom{A}B}A_{BA'} + F_{A'}^{\phantom{A'}B'}A_{AB'}\big) - m^2A_{AA'} = 0 \, .
\label{eq:spin1_EMa_EOM}
\end{align}
It is clear that we can, in principle, generalize the coupling to 
\begin{align}
    \tfrac{1}{2}\big(q F_{A}^{\phantom{A}B}A_{BA'} + \bar{q} F_{A'}^{\phantom{A'}B'}A_{AB'}\big)\,.
\end{align}
On taking the divergence of the equation we arrive at the modified transversality constraint
\begin{align}
D^{AA'}E_{AA'} &= 0 \notag \\
\Leftrightarrow \qquad \frac{q-1}{2}F_{AB}D^{A}_{\phantom{A}B'}A^{BB'} + \frac{q-1}{2}F_{A'B'}D_{B}^{\phantom{B}A'}A^{BB'} - m^2 D^{AA'}A_{AA'} &= 0 \, ,
\label{eq:spin1_EMa_constraint}
\end{align}
where we have already imposed the vacuum equations of motion on the background electromagnetic field. It is not the simple transversality constraint in general, except if $q=1$. However, it is still a constraint in the sense that it fixes one of the four degrees of freedom of the field $A_{AA'}$.\footnote{From the Hamiltonian point of view $\chi_1=p_0$ is still the primary constraint because the interaction is non-derivative ($F^{AA'BB'}$ is a background). There is also a secondary constraint. Therefore, unlocking $q$ has not altered the number of constraints. From the Stueckelberg point of view we see that $F^{\mu\nu}(A_\mu-\pl_\mu\phi )(A_\nu-\pl_\nu \phi)$ does not produce any higher derivatives for $\phi$, only $F^{\mu\nu} A_\mu \pl_\nu \phi$. Therefore, there are no reasons to forbid it. } 

Let us check if it is possible to chiralize the theory for generic $q$. We define the following chiral field
\begin{align}
\varphi_{AB} := m^{-1}D_{(A|A'|}A_{B)}^{\phantom{B)}A'} \, .
\label{eq:spin1_EMa_def_(2,0)}
\end{align}
By using this definition and the Fierz identity \eqref{Fierz}, we can rewrite the second-order equation of motion \eqref{eq:spin1_EMa_EOM} as the first-order one
\begin{align}
2D^{B}_{\phantom{B}A'}\varphi_{AB} + \frac{q+1}{2}F_{A}^{\phantom{A}B}A_{BA'} + \frac{q-1}{2}F_{A'}^{\phantom{A'}B'}A_{AB'} - m^2 A_{AA'} = 0 \, ,
\label{eq:spin1_EMa_EOM_1st_order}
\end{align}
which itself can be rewritten as
\begin{align}
A_{AA'} = 2m^{-1}M_{A\phantom{B}A'}^{\phantom{A}B\phantom{A'}B'}D^{C}_{\phantom{C}B'}\varphi_{BC} \, .
\label{eq:spin1_EMa_def_(1,1)}
\end{align}
Here, the matrix $M$ is defined such that
\begin{align}
\Big(\epsilon_{A}^{\phantom{A}C}\epsilon_{A'}^{\phantom{A'}C'} - \frac{q+1}{2m^2}F_{A}^{\phantom{A}C}\epsilon_{A'}^{\phantom{A'}C'} - \frac{q-1}{2m^2}F_{A'}^{\phantom{A'}C'}\epsilon_{A}^{\phantom{A}C}\Big)M_{C\phantom{B}C'}^{\phantom{C}B\phantom{C'}B'} = \delta_{A}^{B}\delta_{A'}^{B'} \, .
\label{eq:spin1_EMa_def_M}
\end{align}
If we replace $A_{AA'}$ in \eqref{eq:spin1_EMa_def_(2,0)} by its definition \eqref{eq:spin1_EMa_def_(1,1)} in terms of $\varphi_{AB}$, we obtain the following second-order chiral equation of motion
\begin{align}
2D_{(A}^{\phantom{(A}A'}\Big(M_{B)\phantom{C}A'}^{\phantom{B)}C\phantom{A'}B'}D^{D}_{\phantom{D}B'}\varphi_{CD}\Big) + m^2\varphi_{AB} = 0 \, .
\label{eq:spin1_EMa_chiral_EOM}
\end{align}
It can be obtained from the following Lagrangian density
\begin{align}
\mathcal{L} = D_{A}^{\phantom{A}A'}\varphi^{AB}M_{B\phantom{C}A'}^{\phantom{B}C\phantom{A'}B'}D^{D}_{\phantom{D}B'}\varphi_{CD} + \frac{1}{2}m^2\varphi^{AB}\varphi_{AB}\, .
\label{eq:spin1_EMa_Lagrangian_chiral}
\end{align}
This chiral Lagrangian density describes the three degrees of freedom of a massive spin-one field interacting with a vacuum electromagnetic field.

\paragraph{What happened to the transversality constraints?} Let us check if the constraint \eqref{eq:spin1_EMa_constraint} is trivial in the chiral language. In order to do this, let us use the expression of the $(1,1)$-field in terms of the $(2,0)$-field \eqref{eq:spin1_EMa_def_(1,1)} to rewrite the constraint in terms of the chiral field. It becomes
\begin{align}
\frac{q-1}{2m^2}F_{A}^{\phantom{A}B}D^{AA'}\big(M_{B\phantom{C}A'}^{\phantom{B}C\phantom{A'}C'}D^{D}_{\phantom{D}C'}\varphi_{CD}\big) &+ \frac{q-1}{2m^2}F_{A'}^{\phantom{A'}B'}D^{AA'}\big(M_{A\phantom{C}B'}^{\phantom{A}C\phantom{B'}C'}D^{D}_{\phantom{D}C'}\varphi_{CD}\big) \notag \\
&- D_{AA'}\big(M_{A\phantom{C}A'}^{\phantom{A}C\phantom{A'}C'}D^{D}_{\phantom{D}C'}\varphi_{CD}\big) = 0 \, .
\end{align}
We rewrite the definition of $M$ \eqref{eq:spin1_EMa_def_M} as
\begin{align}
M_{A\phantom{C}A'}^{\phantom{A}C\phantom{A'}C'} = \delta_A^C\delta_{A'}^{C'} + \frac{q+1}{2m^2}F_{A}^{\phantom{A}B}M_{B\phantom{C}A'}^{\phantom{B}C\phantom{A'}C'} + \frac{q-1}{2m^2}F_{A'}^{\phantom{A'}B'}M_{A\phantom{C}B'}^{\phantom{A}C\phantom{B'}C'} \, ,
\end{align}
to replace $M$ in the last term of the constraint. The result is
\begin{align}
F^{AB}\Big(2D_{A}^{\phantom{A}A'}\big(M_{B\phantom{C}A'}^{\phantom{B}C\phantom{A'}B'}D^{D}_{\phantom{D}B'}\varphi_{CD}\big) + m^2\varphi_{AB}\Big) = 0 \, ,
\end{align}
where we used the fact that the electromagnetic field satisfies the vacuum equations of motion. This expression is trivially satisfied because of the chiral equation of motion \eqref{eq:spin1_EMa_chiral_EOM}. Note that this result does not depend on the value of the coefficient $q$. It means that the chiralization is achievable for any value of $q$. Note that in the case where $q=0$, we do not need to impose a vacuum electromagnetic field to obtain the constraint \eqref{eq:spin1_EMa_constraint}, but it is still needed if we want to trivialize the constraint in the chiral formulation.

\footnotesize
\providecommand{\href}[2]{#2}\begingroup\raggedright\endgroup


\begin{thebibliography}{10}

\bibitem{Wigner:1939cj}
E.~P. Wigner, ``{On Unitary Representations of the Inhomogeneous Lorentz Group},'' \href{http://dx.doi.org/10.2307/1968551}{{\em Annals Math.} {\bfseries 40} (1939) 149--204}.
[Reprint: Nucl. Phys. Proc. Suppl.6,9(1989)].

\bibitem{Bekaert:2006py}
X.~Bekaert and N.~Boulanger, ``The unitary representations of the poincare group in any spacetime dimension,''
\href{http://arxiv.org/abs/hep-th/0611263}{{\ttfamily hep-th/0611263}}.

\bibitem{Basile:2016aen}
T.~Basile, X.~Bekaert, and N.~Boulanger, ``{Mixed-symmetry fields in de Sitter space: a group theoretical glance},'' \href{http://dx.doi.org/10.1007/JHEP05(2017)081}{{\em JHEP} {\bfseries 05} (2017) 081}, \href{http://arxiv.org/abs/1612.08166}{{\ttfamily arXiv:1612.08166 [hep-th]}}.

\bibitem{Bekaert:2017khg}
X.~Bekaert and E.~D. Skvortsov, ``{Elementary particles with continuous spin},'' \href{http://dx.doi.org/10.1142/S0217751X17300198}{{\em Int. J. Mod. Phys. A} {\bfseries 32} no.~23n24, (2017) 1730019}, \href{http://arxiv.org/abs/1708.01030}{{\ttfamily arXiv:1708.01030 [hep-th]}}.

\bibitem{Bekaert:2022poo}
X.~Bekaert, N.~Boulanger, A.~Campoleoni, M.~Chiodaroli, D.~Francia, M.~Grigoriev, E.~Sezgin, and E.~Skvortsov, ``{Snowmass White Paper: Higher Spin Gravity and Higher Spin Symmetry},'' \href{http://arxiv.org/abs/2205.01567}{{\ttfamily arXiv:2205.01567 [hep-th]}}.

\bibitem{Metsaev:1991mt}
R.~R. Metsaev, ``{Poincare invariant dynamics of massless higher spins: Fourth order analysis on mass shell},''
{\em Mod. Phys. Lett.} {\bfseries A6} (1991) 359--367.

\bibitem{Metsaev:1991nb}
R.~R. Metsaev, ``{$S$ matrix approach to massless higher spins theory. 2: The Case of internal symmetry},''
{\em Mod. Phys. Lett.} {\bfseries A6} (1991) 2411--2421.

\bibitem{Ponomarev:2016lrm}
D.~Ponomarev and E.~D. Skvortsov, ``{Light-Front Higher-Spin Theories in Flat Space},'' {\em J. Phys.} {\bfseries A50} no.~9, (2017) 095401,
\href{http://arxiv.org/abs/1609.04655}{{\ttfamily arXiv:1609.04655 [hep-th]}}.

\bibitem{Skvortsov:2018jea}
E.~D. Skvortsov, T.~Tran, and M.~Tsulaia, ``{Quantum Chiral Higher Spin Gravity},'' {\em Phys. Rev. Lett.} {\bfseries 121} no.~3, (2018) 031601,
\href{http://arxiv.org/abs/1805.00048}{{\ttfamily arXiv:1805.00048 [hep-th]}}.

\bibitem{Skvortsov:2020wtf}
E.~Skvortsov, T.~Tran, and M.~Tsulaia, ``{More on Quantum Chiral Higher Spin Gravity},'' {\em Phys. Rev.} {\bfseries D101} no.~10, (2020) 106001,
\href{http://arxiv.org/abs/2002.08487}{{\ttfamily arXiv:2002.08487 [hep-th]}}.

\bibitem{Ponomarev:2017nrr}
D.~Ponomarev, ``{Chiral Higher Spin Theories and Self-Duality},'' {\em JHEP} {\bfseries 12} (2017) 141,
\href{http://arxiv.org/abs/1710.00270}{{\ttfamily arXiv:1710.00270 [hep-th]}}.

\bibitem{Krasnov:2021nsq}
K.~Krasnov, E.~Skvortsov, and T.~Tran, ``{Actions for Self-dual Higher Spin Gravities},''
\href{http://arxiv.org/abs/2105.12782}{{\ttfamily arXiv:2105.12782 [hep-th]}}.

\bibitem{Buonanno:2022pgc}
A.~Buonanno, M.~Khalil, D.~O'Connell, R.~Roiban, M.~P. Solon, and M.~Zeng, ``{Snowmass White Paper: Gravitational Waves and Scattering Amplitudes},'' in {\em {Snowmass 2021}}.
\newblock 4, 2022.
\newblock \href{http://arxiv.org/abs/2204.05194}{{\ttfamily arXiv:2204.05194 [hep-th]}}.

\bibitem{Francia:2013sca}
D.~Francia, S.~L. Lyakhovich, and A.~A. Sharapov, ``{On the gauge symmetries of Maxwell-like higher-spin Lagrangians},'' \href{http://dx.doi.org/10.1016/j.nuclphysb.2014.02.001}{{\em Nucl. Phys. B} {\bfseries 881} (2014) 248--268}, \href{http://arxiv.org/abs/1310.8589}{{\ttfamily arXiv:1310.8589 [hep-th]}}.

\bibitem{Abakumova:2020ajc}
V.~A. Abakumova and S.~L. Lyakhovich, ``{Hamiltonian constraints and unfree gauge symmetry},'' \href{http://dx.doi.org/10.1103/PhysRevD.102.125003}{{\em Phys. Rev. D} {\bfseries 102} no.~12, (2020) 125003}, \href{http://arxiv.org/abs/2009.02848}{{\ttfamily arXiv:2009.02848 [hep-th]}}.

\bibitem{Fierz:1939ix}
M.~Fierz and W.~Pauli, ``{On relativistic wave equations for particles of arbitrary spin in an electromagnetic field},'' \href{http://dx.doi.org/10.1098/rspa.1939.0140}{{\em Proc. Roy. Soc. Lond. A} {\bfseries 173} (1939) 211--232}.

\bibitem{Singh:1974qz}
L.~P.~S. Singh and C.~R. Hagen, ``{Lagrangian formulation for arbitrary spin. 1. The boson case},''
\href{http://dx.doi.org/10.1103/PhysRevD.9.898}{{\em Phys. Rev.} {\bfseries D9} (1974) 898--909}.

\bibitem{Singh:1974rc}
L.~P.~S. Singh and C.~R. Hagen, ``{Lagrangian formulation for arbitrary spin. 2. The fermion case},''
\href{http://dx.doi.org/10.1103/PhysRevD.9.910}{{\em Phys. Rev.} {\bfseries D9} (1974) 910--920}.

\bibitem{Boulware:1972yco}
D.~G. Boulware and S.~Deser, ``{Can gravitation have a finite range?},'' \href{http://dx.doi.org/10.1103/PhysRevD.6.3368}{{\em Phys. Rev. D} {\bfseries 6} (1972) 3368--3382}.

\bibitem{Bergshoeff:2009hq}
E.~A. Bergshoeff, O.~Hohm, and P.~K. Townsend, ``{Massive Gravity in Three Dimensions},'' \href{http://dx.doi.org/10.1103/PhysRevLett.102.201301}{{\em Phys. Rev. Lett.} {\bfseries 102} (2009) 201301}, \href{http://arxiv.org/abs/0901.1766}{{\ttfamily arXiv:0901.1766 [hep-th]}}.

\bibitem{deRham:2010kj}
C.~de~Rham, G.~Gabadadze, and A.~J. Tolley, ``{Resummation of Massive Gravity},'' \href{http://dx.doi.org/10.1103/PhysRevLett.106.231101}{{\em Phys. Rev. Lett.} {\bfseries 106} (2011) 231101}, \href{http://arxiv.org/abs/1011.1232}{{\ttfamily arXiv:1011.1232 [hep-th]}}.

\bibitem{Hassan:2011zd}
S.~F. Hassan and R.~A. Rosen, ``{Bimetric Gravity from Ghost-free Massive Gravity},'' \href{http://dx.doi.org/10.1007/JHEP02(2012)126}{{\em JHEP} {\bfseries 02} (2012) 126}, \href{http://arxiv.org/abs/1109.3515}{{\ttfamily arXiv:1109.3515 [hep-th]}}.

\bibitem{deRham:2014zqa}
C.~de~Rham, ``{Massive Gravity},'' \href{http://dx.doi.org/10.12942/lrr-2014-7}{{\em Living Rev. Rel.} {\bfseries 17} (2014) 7}, \href{http://arxiv.org/abs/1401.4173}{{\ttfamily arXiv:1401.4173 [hep-th]}}.

\bibitem{Zinoviev:2001dt}
Y.~M. Zinoviev, ``{On massive high spin particles in AdS},'' \href{http://arxiv.org/abs/hep-th/0108192}{{\ttfamily arXiv:hep-th/0108192}}.

\bibitem{Fronsdal:1978rb}
C.~Fronsdal, ``{Massless Fields with Integer Spin},'' \href{http://dx.doi.org/10.1103/PhysRevD.18.3624}{{\em Phys. Rev. D} {\bfseries 18} (1978) 3624}.

\bibitem{Zinoviev:2006im}
Y.~M. Zinoviev, ``{On massive spin 2 interactions},'' \href{http://dx.doi.org/10.1016/j.nuclphysb.2007.02.005}{{\em Nucl. Phys. B} {\bfseries 770} (2007) 83--106}, \href{http://arxiv.org/abs/hep-th/0609170}{{\ttfamily arXiv:hep-th/0609170}}.

\bibitem{Zinoviev:2008ck}
Y.~M. Zinoviev, ``{On spin 3 interacting with gravity},'' \href{http://dx.doi.org/10.1088/0264-9381/26/3/035022}{{\em Class. Quant. Grav.} {\bfseries 26} (2009) 035022}, \href{http://arxiv.org/abs/0805.2226}{{\ttfamily arXiv:0805.2226 [hep-th]}}.

\bibitem{Zinoviev:2009hu}
Y.~M. Zinoviev, ``{On massive spin 2 electromagnetic interactions},'' \href{http://dx.doi.org/10.1016/j.nuclphysb.2009.04.027}{{\em Nucl. Phys. B} {\bfseries 821} (2009) 431--451}, \href{http://arxiv.org/abs/0901.3462}{{\ttfamily arXiv:0901.3462 [hep-th]}}.

\bibitem{Zinoviev:2010cr}
Y.~M. Zinoviev, ``{Spin 3 cubic vertices in a frame-like formalism},'' \href{http://dx.doi.org/10.1007/JHEP08(2010)084}{{\em JHEP} {\bfseries 08} (2010) 084}, \href{http://arxiv.org/abs/1007.0158}{{\ttfamily arXiv:1007.0158 [hep-th]}}.

\bibitem{Buchbinder:2012iz}
I.~L. Buchbinder, T.~V. Snegirev, and Y.~M. Zinoviev, ``{Cubic interaction vertex of higher-spin fields with external electromagnetic field},'' \href{http://dx.doi.org/10.1016/j.nuclphysb.2012.07.012}{{\em Nucl. Phys. B} {\bfseries 864} (2012) 694--721}, \href{http://arxiv.org/abs/1204.2341}{{\ttfamily arXiv:1204.2341 [hep-th]}}.

\bibitem{Pashnev:1989gm}
A.~I. Pashnev, ``{Composite Systems and Field Theory for a Free Regge Trajectory},'' \href{http://dx.doi.org/10.1007/BF01017664}{{\em Theor. Math. Phys.} {\bfseries 78} (1989) 272--277}.

\bibitem{Buchbinder:2005ua}
I.~L. Buchbinder and V.~A. Krykhtin, ``{Gauge invariant Lagrangian construction for massive bosonic higher spin fields in D dimensions},'' \href{http://dx.doi.org/10.1016/j.nuclphysb.2005.07.035}{{\em Nucl. Phys. B} {\bfseries 727} (2005) 537--563}, \href{http://arxiv.org/abs/hep-th/0505092}{{\ttfamily arXiv:hep-th/0505092}}.

\bibitem{Bekaert:2003uc}
X.~Bekaert, I.~L. Buchbinder, A.~Pashnev, and M.~Tsulaia, ``{On higher spin theory: Strings, BRST, dimensional reductions},'' \href{http://dx.doi.org/10.1088/0264-9381/21/10/018}{{\em Class. Quant. Grav.} {\bfseries 21} (2004) S1457--1464}, \href{http://arxiv.org/abs/hep-th/0312252}{{\ttfamily arXiv:hep-th/0312252}}.

\bibitem{Francia:2010qp}
D.~Francia, ``{String theory triplets and higher-spin curvatures},'' {\em Phys.Lett.} {\bfseries B690} (2010) 90--95,
\href{http://arxiv.org/abs/1001.5003}{{\ttfamily arXiv:1001.5003 [hep-th]}}.

\bibitem{Buchbinder:2007ix}
I.~L. Buchbinder, V.~A. Krykhtin, and H.~Takata, ``{Gauge invariant Lagrangian construction for massive bosonic mixed symmetry higher spin fields},'' \href{http://dx.doi.org/10.1016/j.physletb.2007.09.033}{{\em Phys. Lett. B} {\bfseries 656} (2007) 253--264}, \href{http://arxiv.org/abs/0707.2181}{{\ttfamily arXiv:0707.2181 [hep-th]}}.

\bibitem{Buchbinder:2008ss}
I.~L. Buchbinder and A.~V. Galajinsky, ``{Quartet unconstrained formulation for massive higher spin fields},'' \href{http://dx.doi.org/10.1088/1126-6708/2008/11/081}{{\em JHEP} {\bfseries 11} (2008) 081}, \href{http://arxiv.org/abs/0810.2852}{{\ttfamily arXiv:0810.2852 [hep-th]}}.

\bibitem{Kaparulin:2012px}
D.~S. Kaparulin, S.~L. Lyakhovich, and A.~A. Sharapov, ``{Consistent interactions and involution},'' \href{http://dx.doi.org/10.1007/JHEP01(2013)097}{{\em JHEP} {\bfseries 01} (2013) 097}, \href{http://arxiv.org/abs/1210.6821}{{\ttfamily arXiv:1210.6821 [hep-th]}}.

\bibitem{Kazinski:2005eb}
P.~O. Kazinski, S.~L. Lyakhovich, and A.~A. Sharapov, ``{Lagrange structure and quantization},'' \href{http://dx.doi.org/10.1088/1126-6708/2005/07/076}{{\em JHEP} {\bfseries 07} (2005) 076}, \href{http://arxiv.org/abs/hep-th/0506093}{{\ttfamily arXiv:hep-th/0506093}}.

\bibitem{Metsaev:2012uy}
R.~R. Metsaev, ``{BRST-BV approach to cubic interaction vertices for massive and massless higher-spin fields},'' \href{http://dx.doi.org/10.1016/j.physletb.2013.02.009}{{\em Phys. Lett. B} {\bfseries 720} (2013) 237--243}, \href{http://arxiv.org/abs/1205.3131}{{\ttfamily arXiv:1205.3131 [hep-th]}}.

\bibitem{Abakumova:2021evc}
V.~A. Abakumova and S.~L. Lyakhovich, ``{Reducible Stueckelberg symmetry and dualities},'' \href{http://dx.doi.org/10.1016/j.physletb.2021.136552}{{\em Phys. Lett. B} {\bfseries 820} (2021) 136552}, \href{http://arxiv.org/abs/2106.09355}{{\ttfamily arXiv:2106.09355 [hep-th]}}.

\bibitem{Abakumova:2023wve}
V.~A. Abakumova and S.~L. Lyakhovich, ``{Dualisation of free fields},'' \href{http://dx.doi.org/10.1016/j.aop.2023.169322}{{\em Annals Phys.} {\bfseries 453} (2023) 169322}, \href{http://arxiv.org/abs/2303.02616}{{\ttfamily arXiv:2303.02616 [hep-th]}}.

\bibitem{Skvortsov:2023jbn}
E.~Skvortsov and M.~Tsulaia, ``{Cubic action for Spinning Black Holes from massive higher-spin gauge symmetry},'' \href{http://arxiv.org/abs/2312.08184}{{\ttfamily arXiv:2312.08184 [hep-th]}}.

\bibitem{Metsaev:2005ar}
R.~R. Metsaev, ``{Cubic interaction vertices of massive and massless higher spin fields},'' \href{http://dx.doi.org/10.1016/j.nuclphysb.2006.10.002}{{\em Nucl. Phys. B} {\bfseries 759} (2006) 147--201}, \href{http://arxiv.org/abs/hep-th/0512342}{{\ttfamily arXiv:hep-th/0512342}}.

\bibitem{Metsaev:2007rn}
R.~R. Metsaev, ``{Cubic interaction vertices for fermionic and bosonic arbitrary spin fields},'' \href{http://dx.doi.org/10.1016/j.nuclphysb.2012.01.022}{{\em Nucl. Phys. B} {\bfseries 859} (2012) 13--69}, \href{http://arxiv.org/abs/0712.3526}{{\ttfamily arXiv:0712.3526 [hep-th]}}.

\bibitem{Metsaev:2022yvb}
R.~R. Metsaev, ``{Interacting massive and massless arbitrary spin fields in 4d flat space},'' \href{http://arxiv.org/abs/2206.13268}{{\ttfamily arXiv:2206.13268 [hep-th]}}.

\bibitem{Ochirov:2022nqz}
A.~Ochirov and E.~Skvortsov, ``{Chiral Approach to Massive Higher Spins},'' \href{http://dx.doi.org/10.1103/PhysRevLett.129.241601}{{\em Phys. Rev. Lett.} {\bfseries 129} no.~24, (2022) 241601}, \href{http://arxiv.org/abs/2207.14597}{{\ttfamily arXiv:2207.14597 [hep-th]}}.

\bibitem{Chalmers:1997ui}
G.~Chalmers and W.~Siegel, ``{Simplifying algebra in Feynman graphs, Part I: Spinors},'' \href{http://dx.doi.org/10.1103/PhysRevD.59.045012}{{\em Phys. Rev.} {\bfseries D59} (1999) 045012},
\href{http://arxiv.org/abs/hep-ph/9708251}{{\ttfamily arXiv:hep-ph/9708251 [hep-ph]}}.

\bibitem{Chalmers:2001cy}
G.~Chalmers and W.~Siegel, ``{Simplifying algebra in Feynman graphs. 3. Massive vectors},'' \href{http://dx.doi.org/10.1103/PhysRevD.63.125027}{{\em Phys. Rev.} {\bfseries D63} (2001) 125027},
\href{http://arxiv.org/abs/hep-th/0101025}{{\ttfamily arXiv:hep-th/0101025 [hep-th]}}.

\bibitem{Fradkin:1984ai}
E.~S. Fradkin and A.~A. Tseytlin, ``{Quantum Equivalence of Dual Field Theories},'' \href{http://dx.doi.org/10.1016/0003-4916(85)90225-8}{{\em Annals Phys.} {\bfseries 162} (1985) 31}.

\bibitem{Cangemi:2023ysz}
L.~Cangemi, M.~Chiodaroli, H.~Johansson, A.~Ochirov, P.~Pichini, and E.~Skvortsov, ``{From higher-spin gauge interactions to Compton amplitudes for root-Kerr},'' \href{http://dx.doi.org/10.1007/JHEP09(2024)196}{{\em JHEP} {\bfseries 09} (2024) 196}, \href{http://arxiv.org/abs/2311.14668}{{\ttfamily arXiv:2311.14668 [hep-th]}}.

\bibitem{Grigoriev:2016wmk}
M.~Grigoriev, ``{Presymplectic structures and intrinsic Lagrangians},'' \href{http://arxiv.org/abs/1606.07532}{{\ttfamily arXiv:1606.07532 [hep-th]}}.

\bibitem{Grigoriev:2021wgw}
M.~Grigoriev and V.~Gritzaenko, ``{Presymplectic structures and intrinsic Lagrangians for massive fields},'' \href{http://dx.doi.org/10.1016/j.nuclphysb.2022.115686}{{\em Nucl. Phys. B} {\bfseries 975} (2022) 115686}, \href{http://arxiv.org/abs/2109.05596}{{\ttfamily arXiv:2109.05596 [hep-th]}}.

\bibitem{Delfino:2012zy}
G.~Delfino, K.~Krasnov, and C.~Scarinci, ``{Pure Connection Formalism for Gravity: Linearized Theory},'' \href{http://dx.doi.org/10.1007/JHEP03(2015)118}{{\em JHEP} {\bfseries 03} (2015) 118}, \href{http://arxiv.org/abs/1205.7045}{{\ttfamily arXiv:1205.7045 [hep-th]}}.

\bibitem{Arkani-Hamed:2017jhn}
N.~Arkani-Hamed, T.-C. Huang, and Y.-t. Huang, ``{Scattering amplitudes for all masses and spins},'' \href{http://dx.doi.org/10.1007/JHEP11(2021)070}{{\em JHEP} {\bfseries 11} (2021) 070}, \href{http://arxiv.org/abs/1709.04891}{{\ttfamily arXiv:1709.04891 [hep-th]}}.

\bibitem{Guevara:2018wpp}
A.~Guevara, A.~Ochirov, and J.~Vines, ``{Scattering of Spinning Black Holes from Exponentiated Soft Factors},'' \href{http://dx.doi.org/10.1007/JHEP09(2019)056}{{\em JHEP} {\bfseries 09} (2019) 056},
\href{http://arxiv.org/abs/1812.06895}{{\ttfamily arXiv:1812.06895 [hep-th]}}.

\bibitem{Chung:2018kqs}
M.-Z. Chung, Y.-T. Huang, J.-W. Kim, and S.~Lee, ``{The simplest massive S-matrix: from minimal coupling to Black Holes},'' \href{http://dx.doi.org/10.1007/JHEP04(2019)156}{{\em JHEP} {\bfseries 04} (2019) 156},
\href{http://arxiv.org/abs/1812.08752}{{\ttfamily arXiv:1812.08752 [hep-th]}}.

\bibitem{Cangemi:2022bew}
L.~Cangemi, M.~Chiodaroli, H.~Johansson, A.~Ochirov, P.~Pichini, and E.~Skvortsov, ``{Kerr Black Holes From Massive Higher-Spin Gauge Symmetry},'' \href{http://dx.doi.org/10.1103/PhysRevLett.131.221401}{{\em Phys. Rev. Lett.} {\bfseries 131} no.~22, (2023) 221401}, \href{http://arxiv.org/abs/2212.06120}{{\ttfamily arXiv:2212.06120 [hep-th]}}.

\bibitem{Cangemi:2023bpe}
L.~Cangemi, M.~Chiodaroli, H.~Johansson, A.~Ochirov, P.~Pichini, and E.~Skvortsov, ``{Compton Amplitude for Rotating Black Hole from QFT},'' \href{http://dx.doi.org/10.1103/PhysRevLett.133.071601}{{\em Phys. Rev. Lett.} {\bfseries 133} no.~7, (2024) 071601}, \href{http://arxiv.org/abs/2312.14913}{{\ttfamily arXiv:2312.14913 [hep-th]}}.

\end{thebibliography}
\end{document}